\documentclass[runningheads]{llncs}
\usepackage[utf8]{inputenc}
\usepackage[english]{babel}
\let\proof\relax
 
\usepackage{amsthm}
\usepackage{mathtools}
\usepackage{float}
\usepackage{tikz}
\tikzstyle{vertex}=[circle, draw, inner sep=0pt, minimum size=5pt]
\newcommand{\vertex}{\node[vertex]}
\usetikzlibrary{decorations.markings}
\usetikzlibrary{decorations.pathreplacing}
\usetikzlibrary{arrows.meta}
\usepackage{anyfontsize}
\usepackage{graphicx}
\usepackage{amsmath}
\usepackage{amsfonts}
\usepackage{xcolor}
\usepackage{amssymb}
\usepackage{amsmath}
\usetikzlibrary{positioning}
\newcommand{\boundellipse}[3]% center, xdim, ydim
{(#1) ellipse (#2 and #3)
}
\usetikzlibrary{shapes.geometric}
\tikzstyle{square}=[draw, shape=regular polygon, regular polygon sides=4,draw,inner sep=0pt,minimum
size=0.225cm]
\tikzstyle{triangle}=[draw, shape=regular polygon, regular polygon sides=3,draw,inner sep=0pt,minimum
size=0.3cm]
\definecolor{magenta}{rgb}{0.8, 0.0, 0.8}
\definecolor{cyan}{rgb}{0.0, 1.0, 1.0}
\definecolor{green1}{rgb}{0, 1, 0}
\definecolor{green}{rgb}{0, 1, 0}
\definecolor{brown}{rgb}{0.65, 0.16, 0.16}
\definecolor{aquamarine}{rgb}{0.5, 1.0, 0.83}
\definecolor{battleshipgrey}{rgb}{0.52, 0.52, 0.51}
\begin{document}
\title{Offensive Alliances in Graphs}
%
%\titlerunning{Abbreviated paper title}
% If the paper title is too long for the running head, you can set
% an abbreviated paper title here
%
\author{Ajinkya Gaikwad\thanks{The first author  gratefully acknowledges support from the Ministry of Human Resource Development, 
 Government of India, under Prime Minister's Research Fellowship Scheme (No. MRF-192002-211).} \and Soumen Maity\thanks{The 
 second author's research was supported in part by the Science and Engineering Research Board (SERB), Govt. of India, under Sanction Order No. MTR/2018/001025.}}
\authorrunning{A.\,Gaikwad and S.\,Maity}
% First names are abbreviated in the running head.
% If there are more than two authors, 'et al.' is used.
%
\institute{Indian Institute of Science Education and Research, Pune, India 
\email{\texttt{ajinkya.gaikwad@students.iiserpune.ac.in;}}
\email{\texttt{soumen@iiserpune.ac.in}}
}
\maketitle              % typeset the header of the contribution
\begin{abstract}  The {\sc Offensive Alliance} problem has been studied extensively during the last twenty years. A set $S\subseteq V$ of vertices is an offensive alliance  in an undirected graph $G=(V,E)$ if
 each  $v\in N(S)$ has at least as many neighbours in $S$ as it has neighbours (including itself)
 not in $S$.   
We study the classical and parameterized complexity of the 
{\sc Offensive Alliance} problem, where the aim is to find a minimum size
offensive alliance.
%Our focus here lies on natural parameter as well as parameters that measure the structural properties of the input instance. 
We enhance our understanding of the problem from the viewpoint of parameterized 
complexity   by showing that (1)
the problem is W[1]-hard parameterized by a wide range of 
fairly restrictive structural parameters such as the feedback vertex set number, treewidth, pathwidth, and treedepth of the input graph; 
we thereby resolve an open question stated by Bernhard Bliem  and Stefan Woltran  (2018) concerning the complexity of 
{\sc Offensive Alliance} parameterized by treewidth,
(2) unless ETH fails, {\sc Offensive Alliance} problem cannot be solved in time $\mathcal{O}^{*}(2^{o(k \log k)})$ where $k$ is the solution size,
(3) {\sc Offensive Alliance} problem does not admit a polynomial kernel parameterized by solution size and vertex cover of the input graph.
On the positive side we prove that
(4) {\sc Offensive Alliance} can be solved in time $\mathcal{O}^{*}(\tt{vc(G)}^{\mathcal{O}(\tt{vc(G)})})$ where $\tt{vc(G)}$ is the vertex cover number of the input graph.
%(5) {\sc Offensive Alliance} problem admits an FPT algorithm when parameterized by vertex integrity of input graph.
In terms of classical complexity, we prove that (5) {\sc Offensive Alliance} cannot be solved in time  $2^{o(n)}$ even when restricted to bipartite graphs, unless  ETH fails, (6)  {\sc Offensive Alliance}  cannot be solved in time $2^{o(\sqrt{n})}$  even when restricted to apex graphs, unless  ETH fails. We also prove that  (7)  {\sc Offensive Alliance}  is NP-complete even when restricted to bipartite, chordal, split and circle graphs.

\end{abstract}

\section{Introduction}

This paper studies the {\sc Offensive Alliance} problem: given an
undirected graph $G$ and a positive integer $r$, determine whether $G$ admits an offensive alliance of size at most $r$. 
It is not surprising that the complexity of {\sc Offensive Alliance} and several of its variants has been studied extensively by the theory community in the past years.
The decision version for several types of alliances have been shown to be NP-complete. 
For an integer $\ell$, a nonempty set $S\subseteq V(G)$ is a {\it defensive $\ell$-alliance} if for each 
$v\in S$, $d_S(v)\geq d_{S^c}(v)+\ell$. A set is a defensive alliance if it is a defensive 
$(-1)$-alliance. A defensive $\ell$-alliance $S$ is {\it global} if $S$ is a dominating set. 
 The defensive $\ell$-alliance problem   is NP-complete for any $\ell$ \cite{SIGARRETA20091687}. The defensive alliance problem is 
 NP-complete even when restricted to split, chordal and bipartite graph \cite{Lindsay}. 
 For an integer $\ell$, a nonempty set $S\subseteq V(G)$ is an {\it offensive $\ell$-alliance} if for each 
$v\in N(S)$, $d_S(v)\geq d_{S^c}(v)+\ell$. An offensive 1-alliance is called an offensive
alliance.  An offensive $\ell$-alliance $S$ is {\it global} if $S$ is a dominating set. 
 Fernau et al. showed that the offensive $\ell$-alliance and global 
 offensive $\ell$-alliance problems are NP-complete for any fixed $\ell$ \cite{FERNAU2009177}. 
 They also proved that for $\ell>1$, $\ell$-offensive alliance is NP-hard, even when restricted to 
 $r$-regular planar graphs.  There are polynomial time algorithms for finding minimum alliances
 in trees \cite{CHANG2012479,Lindsay}.  A polynomial time algorithm for finding minimum defensive alliance in series parallel graph is presented in \cite{10.5555/1292785}. 
 \par Fernau  and Raible showed in \cite{Fernau} that the defensive, offensive and 
 powerful alliance problems and their global
 variants are fixed parameter tractable when parameterized by solution size $r$. Kiyomi and Otachi
 showed in 
 \cite{KIYOMI201791}, the problems of finding smallest alliances of all kinds are fixed-parameter tractable
 when parameteried by the vertex cover number. The problems of finding smallest defensive 
 and offensive alliances are also fixed-parameter tractable
 when parameteried by the neighbourhood diversity \cite{ICDCIT2021}. 
 Enciso \cite{Enciso2009AlliancesIG} proved that 
 finding defensive and global defensive alliances is fixed parameter tractable when parameterized by domino treewidth. 
 Bliem and Woltran \cite{BLIEM2018334} proved that 
  deciding if a graph contains a defensive alliance of size at most
$r$  is W[1]-hard when parameterized by treewidth of the input graph. This puts it among the few problems that are FPT when parameterized by solution size but not when parameterized by treewidth (unless FPT=W[1]). \\

\noindent  \emph{Contribution:} The parameterized complexity of {\sc Offensive Alliance}  parameterized by several structural parameters has remained unexplored.
We resolve the problem with most of these parameters.
We  mostly discuss the parameters that deal with sparseness of graph. 
We show that the problem is  W[1]-hard parameterized by any of the following parameters: the feedback vertex set number, treewidth, pathwidth,
and treedepth of the input graph. Interestingly, our result is significantly stronger 
since we show that hardness even applies in the case that the remaining parts, after deleting the feedback vertex set, are trees of height at most seven.
Next, we turn our attention to parameters vertex cover number and solution size. 
As mentioned before, it is already proved that {\sc Offensive Alliance} problem admits  FPT algorithms parameterized by each of these parameters individually. 
As there is no hope to get FPT algorithms with small structural parameters, we need to make the most out of these two parameters by obtaining efficient algorithms and kernels. The algorithm mentioned in  \cite{Enciso2009AlliancesIG}, has a running time $\mathcal{O}^{*}(2^{\mathcal{O}(r \log r)})$. 
The first question that arises from here is that whether we can get a single 
exponential algorithm? 
We answer this question in a negative way by proving that unless ETH fails, 
{\sc Offensive Alliance} problem cannot be solved in time $\mathcal{O}^{*}(2^{o(r \log r)})$.
For the parameter vertex cover number, the algorithm mentioned in \cite{KIYOMI201791} has running time $\mathcal{O}^{*}({(2^{\tt{vc(G)}})}^{\mathcal{O}(2^{\tt{vc(G)}})})$. In this case, we improve the running time  to $\mathcal{O}^{*}(\tt{vc(G)}^{\mathcal{O}(\tt{vc(G)}})$ where $\tt{vc(G)}$ is the vertex cover number of the input graph. 
Finally, we show that it is unlikely to get polynomial kernel when
parameterized by both of these parameters combined. 
%On the positive side, we have showed that the problem admits an FPT algorithm when parameterized by vertex integrity of the input graph.
\par In search of efficient algorithms, alliance problems 
have been studied on special graph classes. There are polynomial time algorithms for finding minimum alliances in trees \cite{CHANG2012479,Lindsay}.  A polynomial time algorithm for finding minimum defensive alliance in series parallel graph is given in \cite{10.5555/1292785}.
But still, alliance problems remained unexplored on special classes of intersection 
graphs such as interval graphs, circle graphs, circular arc graphs, unit disk graphs etc. We show that the problem remains NP-hard even when restricted to  bipartite, chordal, split and circle graphs. We also prove that the known algorithms on general graphs and apex graphs are unlikely to improve. This is done by showing that the problem cannot be solved in $2^{o(n)}$ time  even when restricted to bipartite graphs and also the problem cannot be solved in $2^{o(\sqrt{n})}$ time even when restricted to apex graphs, unless ETH fails.

\section{Preliminaries}
In real life, an alliance is a collection of people, groups, or states such that the union is 
stronger than individual. The alliance can be either to achieve some common purpose, to protect against 
attack, or to assert collective will against others. This motivates the definitions of defensive and offensive
alliances in graphs. 
The properties of alliances in graphs were first studied by Kristiansen, Hedetniemi, and Hedetniemi 
\cite{kris}. 
They introduced defensive, offensive and powerful alliances. An alliance is global  if it is a dominating set. 
The alliance problems have been studied extensively during last twenty years \cite{frick,SIGARRETA20061345,chel,ROD,SIGA}, and generalizations called $r$-alliances are also studied \cite{SIGARRETA20091687}. 
Throughout this article, $G=(V,E)$ denotes a finite, simple and undirected graph of order $|V|=n$. The subgraph induced by 
$S\subseteq V(G)$ is denoted by $G[S]$. For a vertex $v\in V$, we use $N_G(v)=\{u~:~(u,v)\in E(G)\}$ to denote the (open) neighbourhood 
of vertex $v$ in $G$, and $N_G[v]=N_G(v)\cup \{v\}$ to denote the closed neighbourhood of $v$. The degree $d_G(v)$ of a vertex 
$v\in V(G)$ is $|N_G(v)|$. For a subset $S\subseteq V(G)$, we define its closed neighbourhood as $N_G[S]=\bigcup_{v\in S} N_G[v]$ and its 
open neighbourhood as $N_G(S)=N_G[S]\setminus S$. 
%A set  $S\subseteq V $ is a dominating set of $G$ if $N_G[S]=V$.
For a non-empty subset $S\subseteq V$ and a vertex $v\in V(G)$, $N_S(v)$ denotes the set of neighbours of $v$ in $S$, that is, 
$N_S(v)=\{ u\in S~:~ (u,v)\in E(G)\}$.  We use $d_S(v)=|N_S(v)|$ to denote the degree of vertex $v$ in $G[S]$. 
The complement of the vertex set $S$ in $V$ is denoted by $S^c$.

%\begin{definition}\rm
%A non-empty set 
%$S\subseteq V$ is a  defensive alliance in $G=(V,E)$ if 
%$d_S(v)+1\geq d_{S^c}(v)$ for all $v\in S$.
%\end{definition}
%\noindent A vertex $v\in S$ is said to be  protected if $d_S(v)+1\geq d_{S^c}(v)$. A set $S\subseteq V$ is a  defensive alliance if every vertex in $S$ is  protected. 
\begin{definition}\rm
A non-empty set 
$S\subseteq V$ is a  {\it defensive  alliance} in $G$ if 
$d_S(v)+1\geq d_{S^c}(v)$ for all $v\in S$.
\end{definition}
%Using national security issues to illustrate these concepts, 
\noindent Since each vertex in a defensive alliance $S$ has at least as many vertices from 
its closed neighbor in $S$ as it has in $S^c$, by strength of numbers, 
we say that every vertex in $S$ can be {\it defended} from possible attack by vertices 
in $S^c$. 

\begin{definition}\rm
A non-empty set 
$S\subseteq V$ is an offensive alliance in $G$ if 
$d_S(v)\geq d_{S^c}(v)+1$ for all $v\in N(S)$.
\end{definition}
\noindent Since each vertex in $N(S)$ has more neighbors in $S$ than in $S^c$, we say that every vertex in $N(S)$ is {\it vulnerable} to possible attack by vertices in $S$. Equivalently, since an attack by the vertices in $S$ on the vertices in $V \setminus S$ can result in no worse than a “tie” for $S$, we say that $S$ can effectively attack $N(S)$.

\begin{definition}\rm
A non-empty set 
$S\subseteq V$ is a strong offensive alliance in $G$ if 
$d_S(v)\geq d_{S^c}(v)+2$ for all $v\in N(S)$.
\end{definition}

%\noindent A vertex $v\in N(S)$ is said to be  protected by $S$ if $d_S(v)\geq d_{S^c}(v)+1$. 
% A set $S\subseteq V$ is an offensive alliance if every vertex
% in $N(S)$ is  protected.  Informally, given a graph $G=(V,E)$,
%  we say a set $S$ is an offensive alliance if every vertex that is adjacent to $S$ is outgunned
%  by $S$; more of its neighbours are in $S$ than outside $S$. 
 
\noindent In this paper, we consider {\sc Offensive Alliance}
and {\sc Strong Offensive Alliance} problems under structural parameters. We define these problems as follows:  
    \vspace{3mm}
    \\
    \fbox
    {\begin{minipage}{33.7em}\label{OA }
       {\sc  Offensive Alliance}\\
        \noindent{\bf Input:} An undirected graph $G=(V,E)$ and an  integer $r\geq 1$.
    
        \noindent{\bf Question:} Is there an  offensive alliance $S\subseteq V(G)$ such that 
        $1\leq |S|\leq r$?
    \end{minipage} }
    \vspace{3mm}
    \\
    \fbox
    {\begin{minipage}{33.7em}\label{SOA}
       {\sc Strong Offensive Alliance}\\
        \noindent{\bf Input:} An undirected graph $G=(V,E)$ and an  integer $r\geq 1$.
    
        \noindent{\bf Question:} Is there a strong offensive alliance $S\subseteq V(G)$ such that 
        $1\leq |S|\leq r $?
    \end{minipage} }
    \vspace{3mm}

\noindent For standard notations and definitions in graph theory, we refer to West \cite{west}.
     For the standard concepts in parameterized complexity, see the recent textbook by Cygan et al. \cite{marekcygan}.
     The graph parameters we explicitly use in this paper are  feedback vertex set number, pathwidth, treewidth and 
treedepth. 

%\begin{definition} \rm 
% The {\it vertex cover number} is the size of a minimum vertex cover in a graph $G$ and it is denoted by
% $vc(G)$.
%\end{definition}

    \begin{definition} {\rm
        For a graph $G = (V,E)$, the parameter {\it feedback vertex set} is the cardinality of a smallest set $S \subseteq V(G)$ such that the graph $G-S$ is a forest and it is denoted by $fvs(G)$.}
    \end{definition}

\noindent We now review the concept of a tree decomposition, introduced by Robertson and Seymour in \cite{Neil}.
Treewidth is a  measure of how “tree-like” the graph is.

\begin{definition}\rm \cite{Downey} A {\it tree decomposition} of a graph $G=(V,E)$  is a tree $T$ together with a 
collection of subsets $X_t$ (called \emph{bags}) of $V$ labeled by the vertices $t$ of $T$ such that 
$\bigcup_{t\in T}X_t=V $ and (1) and (2) below hold:
\begin{enumerate}
			\item For every edge $uv \in E(G)$, there  is some $t$ such that $\{u,v\}\subseteq X_t$.
			\item  (Interpolation Property) If $t$ is a vertex on the unique path in $T$ from $t_1$ to $t_2$, then 
			$X_{t_1}\cap X_{t_2}\subseteq X_t$.
		\end{enumerate}
	\end{definition}
	
%\noindent It is important to note that a graph may have several different tree decomposition.
%	Similarly, the same tree decomposition can be valid for several different graphs. 
%	Every graph has a trivial tree decomposition for which $T$ has only one vertex including all
%	of $V$. However, this is not effective for the purpose of solving problems. 
\begin{definition}\rm \cite{Downey} The {\it width} of a tree decomposition is
the maximum value of $|X_t|-1 $ taken over all the vertices $t$ of the tree $T$ of the decomposition.
The treewidth  of a graph $G$  is the  minimum width among all possible tree decompositions of $G$.
\end{definition} 
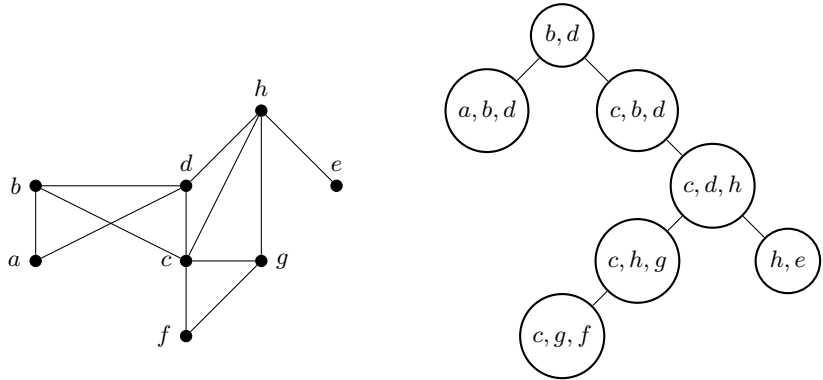
\begin{figure}[ht]
     \centering
 \begin{tikzpicture}[scale=1]
 \node[fill=black, circle, draw=black, inner sep=0, minimum size=0.15cm](b) at (0,3) [label=left: $b$] {};
 \node[fill=black, circle, draw=black, inner sep=0, minimum size=0.15cm](a) at (0,2) [label=left:$a$] {}; 
 \node[fill=black, circle, draw=black, inner sep=0, minimum size=0.15cm](c) at (2,2) [label=left: $c$] {};
 \node[fill=black, circle, draw=black, inner sep=0, minimum size=0.15cm](d) at (2,3) [label=above:$d$] {}; 
 \node[fill=black, circle, draw=black, inner sep=0, minimum size=0.15cm](h) at (3,4) [label=above:$h$] {}; 
 \node[fill=black, circle, draw=black, inner sep=0, minimum size=0.15cm](e) at (4,3) [label=above:$e$] {}; 
 \node[fill=black, circle, draw=black, inner sep=0, minimum size=0.15cm](f) at (2,1) [label=left:$f$] {}; 
 \node[fill=black, circle, draw=black, inner sep=0, minimum size=0.15cm](g) at (3,2) [label=right:$g$] {}; 
 \node[thick, circle,draw, minimum size=.1cm ] (bd) at (7,5) []{{$b,d$}};
% \draw [draw=black, thick](7,5) ellipse (0.6cm and 0.3cm)[]{$bd$};
\node[thick, circle,draw, minimum size=.1cm ] (abd) at (6,4) []{{$a,b,d$}};
\node[thick, circle,draw, minimum size=.1cm ] (cbd) at (8,4) []{{$c,b,d$}};
\node[thick, circle,draw, minimum size=.1cm ] (cdh) at (9,3) []{{$c,d,h$}};
\node[thick, circle,draw, minimum size=.1cm ] (chg) at (8,2) []{{$c,h,g$}};
\node[thick, circle,draw, minimum size=.1cm ] (he) at (10,2) []{{$h,e$}};
\node[thick, circle,draw, minimum size=.1cm ] (cgf) at (7,1) []{{$c,g,f$}};
\path 
(bd) edge (abd)
(bd) edge (cbd)
(cbd) edge (cdh)
(cdh) edge (chg)
(cdh) edge (he)
(chg) edge (cgf);
\path
(a) edge (b)
(b) edge (d)
(b) edge (c)
(a) edge (d)
(d) edge (c)
(d) edge (h)
(h) edge (e)
(h) edge (c)
(c) edge (g)
(f) edge (g)
(h) edge (g)
(c) edge (f);

\end{tikzpicture}
     \caption{Example of a tree decomposition of width 2 }
     \label{fig:treewidth}
 \end{figure}

\begin{example}
Figure \ref{fig:treewidth} gives an example of a tree decomposition of width 2. 
\end{example}

\begin{definition}\rm 
    If the tree $T$ of a tree decomposition is a path, then we say that the tree decomposition 
    is a {\it path decomposition}, and use {\it  pathwidth} in place of treewidth. 
\end{definition}

A rooted forest is a disjoint union of rooted trees. Given a rooted forest $F$, its \emph{transitive closure} is a graph $H$ in which $V(H)$ contains all the nodes of the rooted forest, and $E(H)$ contain an edge between two vertices only if those two vertices form an ancestor-descendant pair in the forest $F$.

   \begin{definition}
        {\rm  The {\it treedepth} of a graph $G$ is the minimum height of a rooted forest $F$ whose transitive closure contains the graph $G$. It is denoted by $td(G)$.}
    \end{definition}
    
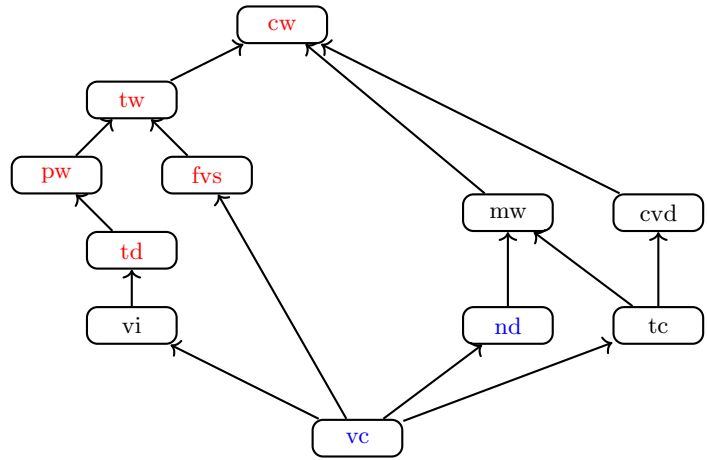
\begin{figure}
     \centering
 \begin{tikzpicture}[%
    auto,
    block/.style={
      rectangle,
      draw=black,
      thick,
      text width=3em,
      align=center,
      rounded corners,
      minimum height=1.5em
    }
]
    \draw (0,0) node[block] (vc) {\color{blue} vc}
          (2,1.5) node[block] (nd) {\color{blue} nd}
          (4,1.5) node[block] (tc) {tc}
          (-3,1.5) node[block] (vi) { vi}
          (-3,2.5) node[block] (td) { \color{red}  td}
          (-2,3.5) node[block] (fvs) {\color{red} fvs}
          (-4,3.5) node[block] (pw) { \color{red} pw}
          (2,3) node[block] (mw) {mw}
          (4,3) node[block] (cvd) {cvd}
          (-3,4.5) node[block] (tw) {\color{red} tw}
          (-1,5.5) node[block] (cw) {\color{red} cw};
          
\draw[->, thick] (vc)--(vi); 
\draw[->, thick] (vc)--(nd); 
\draw[->, thick] (vc)--(tc); 
\draw[->, thick] (vi)--(td); 
\draw[->, thick] (td)--(pw); 
\draw[->, thick] (vc)--(fvs); 
\draw[->, thick] (nd)--(mw); 
\draw[->, thick] (tc)--(mw); 
\draw[->, thick] (tc)--(cvd); 
\draw[->, thick] (pw)--(tw); 
\draw[->, thick] (fvs)--(tw); 
\draw[->, thick] (tw)--(cw); 
\draw[->, thick] (mw)--(cw); 
\draw[->, thick] (cvd)--(cw);

%\draw[green, thick, dashed] (-4,-.5)--(5,-0.5); 
%\draw[green, thick, dashed] (-4,1.8)--(5,1.8); 
%\draw[green, thick, dashed] (-4,1.8)--(-4,-.5); 
%\draw[green, thick, dashed] (5,-.5)--(5,1.8); 

%\draw[red, thick, dashed] (-5,2)--(0,2); 
%\draw[red, thick, dashed] (-5,6)--(0,6); 
%\draw[red, thick, dashed] (-5,2)--(-5,6); 
%\draw[red, thick, dashed] (0,2)--(0,6); 

%\draw[blue, thick, dashed] (1,2)--(5,2); 
%\draw[blue, thick, dashed] (1,6)--(5,6); 
%\draw[blue, thick, dashed] (1,2)--(1,6); 
%\draw[blue, thick, dashed] (5,2)--(5,6); 

  \end{tikzpicture}

\caption{ Relationship between vertex cover (vc), neighbourhood diversity (nd), twin cover (tc), modular width (mw), cluster vertex deletion number (cvd), feedback vertex set (fvs), pathwidth (pw), treewidth (tw) and clique width (cw). 
%Arrow indicate generalizations, for example, treewidth generalizes both feedback vertex set and pathwidth.
Note that $A\rightarrow B$ means that there exists a function $f$ such that for 
all graphs, $f(A(G))\geq B(G)$. It also gives an overview of the parameterized complexity landscape for the {\sc Offensive alliance} problem with general thresholds.
The problem is FPT parameterized by blue colored parameters and W[1]-hard 
when parameterized by red colored parameters. The problem remains unsettled when parameterized by twin cover, modular width and cluster vertex deletion number.}

\label{fig:str}
 \end{figure}
\subsection{Parameterized Complexity}  
A \emph{parameterized problem} is a language $L\subseteq \Sigma^{\star} \times \mathbb{N}$,
where $\Sigma $
is a fixed, finite alphabet. For an instance $(x,k)\in \Sigma^{\star} \times \mathbb{N}$,
$k$ is called the \emph{parameter}. A parameterized problem $\mathcal{P}$ is \emph{fixed-parameter tractable} (FPT in short) if a given instance $(x,k)$ can be solved in time $f(k)\cdot {|(x,k)|}^c$
 where $f$ is some (usually computable) function, and $c$ is a constant. Parameterized complexity classes are defined with respect to {\it fpt-reducibility}. A parameterized problem $\mathcal{P}$ is {\it fpt-reducible} to 
 $\mathcal{Q}$ if in time  $f(k) \cdot {|(x,k)|}^c$, one can transform an instance 
 $(x, k)$ of $\mathcal{P}$ into an instance $(x', k')$ of $Q$ such that $(x, k) \in \mathcal{P}$
 if and only if  $(x',k') \in \mathcal{Q}$, and $k'\leq g(k)$, where $f$ and $g$ are computable functions depending only on $k$. Owing to the definition, if $\mathcal{P}$ {\it fpt-reduces} to $\mathcal{Q}$ and $\mathcal{Q}$ is fixed-parameter tractable then $\mathcal{P}$ is fixed-parameter tractable as well.

 What makes the theory more interesting is a hierarchy of intractable parameterized problem classes above FPT which helps in distinguishing those problems that are not fixed parameter tractable. 
 Central to parameterized complexity is the following hierarchy of complexity classes, defined by the closure of canonical problems under {\it fpt-reductions}: FPT $\subseteq$ W[1] $\subseteq$ W[2] $\subseteq \ldots \subseteq $ XP. All inclusions are believed to be strict. In particular, FPT $\neq$  W[1] under the Exponential Time Hypothesis \cite{Paturi}.
The class W[1] is the analog of NP in parameterized complexity. A major goal in parameterized complexity is to distinguish between parameterized problems which are in FPT
 and those which are \emph{W[1]-hard}, i.e., those to which every problem in W[1] is \emph{fpt-reducible}. There are many problems shown to be complete for W[1], or equivalently \emph{W[1]-complete}, including the {\sc MultiColored Clique} (MCC) problem \cite{Downey}.

 Closely related to fixed-parameter tractability is the notion of preprocessing. A \emph{reduction to a problem kernel}, or equivalently, \emph{problem kernelization} means to apply a data reduction process in polynomial time to an instance $(x, k)$ such that for the reduced instance $(x', k')$ it holds that $(x',k')$ is equivalent to $(x,k)$, $|x'| \leq g(k)$ and $k' \leq g(k)$ for some function $g$ only depending on $k$. Such a reduced instance is called a \emph{problem kernel}. It is easy to show that a parameterized problem is in FPT if and only if there is kernelization algorithm. A \emph{polynomial kernel} is a kernel, whose size can be bounded by a polynomial in the parameter.
 We refer to \cite{marekcygan,Downey} for further details on parameterized complexity. 

\section{W[1]-Hardness Parameterized by Structural Parameters}
In this section we show that {\sc Offensive Alliance} is W[1]-hard parameterized by a vertex deletion set to 
trees of height at most seven, that is, a subset $D$ of the vertices of the graph such that 
every component in the graph, after removing $D$, is a tree of height at most seven. 
On the way towards this result, we provide hardness results for several interesting 
versions of the  {\sc Offensive Alliance} problem which  we require in our proofs. 
 The problem 
 {\sc Offensive Alliance$^{\mbox{F}}$} generalizes {\sc Offensive Alliance} 
 where some vertices are 
 forced to be outside the solution; these vertices are called forbidden vertices.  This variant can be formalized as
follows: 
 \vspace{3mm}
    \\
    \fbox
    {\begin{minipage}{33.7em}\label{OAF}
       {\sc Offensive Alliance$^{\mbox{F}}$}\\
        \noindent{\bf Input:} An undirected graph $G=(V,E)$, an integer $r$ and
        a set $V_{\square}\subseteq V$ of forbidden vertices such that each degree one
        forbidden vertex is adjacent to another forbidden vertex and each  forbidden vertex of degree greater than one is adjacent to a degree one forbidden vertex. \\
        \noindent{\bf Question:} Is there an offensive alliance $S\subseteq V$ such that (i) $1\leq |S|\leq r$, and 
    (ii) $S\cap V_{\square}=\emptyset$?
    %, and (iii) for all $(a,b)\in C$, $V_1$ 
    %contains either $a$ or $b$ but not both?
    \end{minipage} }\\
{\sc  Strong Offensive Alliance$^{\mbox{FN}}$} is a generalization of {\sc  Strong Offensive Alliance$^{\mbox{F}}$} that, in addition, requires
 some ``necessary'' vertices  to be in  $S$.
  This variant can be formalized as
follows: 
\vspace{3mm}
    \\
    \fbox
    {\begin{minipage}{33.7em}\label{OAFN}
       {\sc Strong Offensive Alliance$^{\mbox{FN}}$}\\
        \noindent{\bf Input:} An undirected graph $G=(V,E)$, an integer $r$, a set $V_{\triangle}\subseteq V$, and
        a set $V_{\square}\subseteq V$ of forbidden vertices such that  each degree one
        forbidden vertex is adjacent to another forbidden vertex and each  forbidden vertex of degree greater than one is adjacent to a degree one forbidden vertex.  \\
        %, and 
        %a set $C\subseteq V(G)\times V(G)$.\\
    \noindent{\bf Question:} Is there a strong offensive alliance $S\subseteq V$ such that (i) $1\leq |S|\leq r$, 
    (ii) $S\cap V_{\square}=\emptyset$, and (iii) $V_{\triangle} \subseteq S$?
    %, and (iii) for all $(a,b)\in C$, $V_1$ 
    %contains either $a$ or $b$ but not both?
    \end{minipage} }\\
    
 \noindent While the {\sc  Offensive Alliance}  problem asks for  offensive alliance of 
 size at most $r$,
 we also consider the {\sc Exact  Offensive Alliance} problem that concerns  offensive alliance
 of size exactly $r$. Analogously, we also define exact versions of 
 {\sc Strong Offensive Alliance} presented above. To prove Lemma \ref{oatheorem1}, we consider 
 a variant of the following problem:
 \noindent\vspace{3mm}
    \\
    \fbox
    {\begin{minipage}{33.7em}\label{MSS}
       {\sc  Multidimensional Subset Sum (MSS)}\\
     \noindent{\bf  Input:} An integer $k$, a set 
     $S = \{s_1,\ldots,s_n\}$ of vectors with $s_i \in \mathbb{N}^k$ for every $i$ with 
     $1 \leq i \leq  n$  and a target vector $t \in \mathbb{N}^k$.\\
\noindent {\bf Parameter}: $k$ \\
\noindent{\bf Question}: Is there a subset $S'\subseteq S $ such that $\sum\limits_{s\in S'}{s}=t$?
    \end{minipage} }\\

  \noindent 
  In the {\sc Multidimensional Relaxed Subset Sum} ({\sc MRSS}) problem, an additional integer $k'$ is given
    (which will be part of the parameter)
    and we ask whether there is a subset $S'\subseteq S$ with $|S'|\leq k'$ such that $\sum\limits_{s\in S'}{s}\geq t$.
    More formally,
  % It is known that {\sc MRSS} is W[1]-hard when parameterized by the combined parameter $k+k'$,
   %even if all integers in the input are given in unary \cite{mss}.
\noindent\vspace{3mm}
    \\
    \fbox
    {\begin{minipage}{33.7em}\label{MRSS}
       {\sc Multidimensional Relaxed Subset Sum (MRSS)}\\
     \noindent{\bf  Input:} Two integers $k$ and $k'$, a set 
     $S = \{s_1,\ldots,s_n\}$ of vectors with $s_i \in \mathbb{N}^k$ for every $i$ with 
     $1 \leq i \leq  n$  and a target vector $t \in \mathbb{N}^k$.\\
\noindent {\bf Parameter}: $k+k'$ \\
\noindent{\bf Question}: Is there a subset $S'\subseteq S $ with $|S'|\leq k'$ such that $\sum\limits_{s\in S'}{s}\geq t$?
    \end{minipage} }\\
\begin{lemma}\rm \cite{mss}
{\sc MRSS} is W[1]-hard when parameterized by the combined parameter $k+k'$, even if all integers in the input are given in unary.
\end{lemma}
\noindent We now show that the {\sc Strong Offensive Alliance$^{\mbox{FN}}$} problem is W[1]-hard parameterized by  the size of a vertex 
 deletion set into trees of height at most 5, via a reduction from MRSS.

 \begin{lemma}\label{oatheorem1}\rm
 The {\sc Strong Offensive Alliance$^{\mbox{FN}}$} problem is W[1]-hard when parameterized by  the 
 size of a vertex deletion set into trees of height at most 5.
 \end{lemma}
 
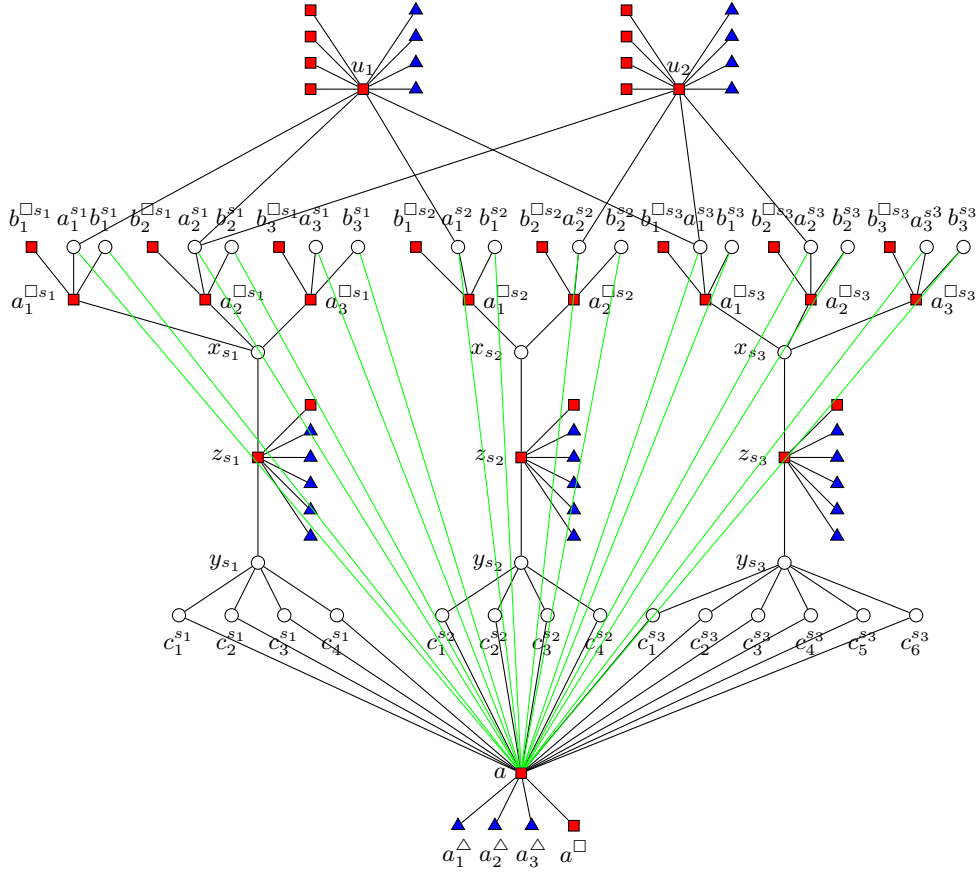
\begin{figure}[ht]
\centering
\begin{tikzpicture}[scale=0.7]
\node[fill=red, square, draw=black, inner sep=0pt, minimum size=0.2cm] (u1) at (-3, 7) [label=above:${u_1}$] {}; 
\node[fill=red, square, draw=black, inner sep=0pt, minimum size=0.2cm] (u11) at (-4, 7) [] {}; 
\node[fill=red, square, draw=black, inner sep=0pt, minimum size=0.2cm] (u12) at (-4, 7.5) [] {}; 
\node[fill=red, square, draw=black, inner sep=0pt, minimum size=0.2cm] (u13) at (-4, 8) [] {}; 
\node[fill=red, square, draw=black, inner sep=0pt, minimum size=0.2cm] (u14) at (-4, 8.5) [] {};

\node[triangle, draw=black, fill= blue, inner sep=0pt, minimum size=0.2cm] (ut11) at (-2,7) [] {};
\node[triangle, draw=black, fill= blue, inner sep=0pt, minimum size=0.2cm] (ut12) at (-2,7.5) [] {};
\node[triangle, draw=black, fill= blue, inner sep=0pt, minimum size=0.2cm] (ut13) at (-2,8) [] {};
\node[triangle, draw=black, fill= blue, inner sep=0pt, minimum size=0.2cm] (ut14) at (-2,8.5) [] {};

\node[fill=red, square, draw=black, inner sep=0pt, minimum size=0.2cm] (u2) at (3, 7) [label=above:${u_2}$] {}; 
\node[fill=red, square, draw=black, inner sep=0pt, minimum size=0.2cm] (u21) at (2, 8) [] {}; 
\node[fill=red, square, draw=black, inner sep=0pt, minimum size=0.2cm] (u22) at (2, 7.5) [] {}; 
\node[fill=red, square, draw=black, inner sep=0pt, minimum size=0.2cm] (u23) at (2, 7) [] {}; 
\node[fill=red, square, draw=black, inner sep=0pt, minimum size=0.2cm] (u24) at (2, 8.5) [] {};

\node[triangle, draw=black, fill= blue, inner sep=0pt, minimum size=0.2cm] (ut1) at (4,7) [] {};
\node[triangle, draw=black, fill= blue, inner sep=0pt, minimum size=0.2cm] (ut2) at (4,7.5) [] {};
\node[triangle, draw=black, fill= blue, inner sep=0pt, minimum size=0.2cm] (ut3) at (4,8) [] {};
\node[triangle, draw=black, fill= blue, inner sep=0pt, minimum size=0.2cm] (ut4) at (4,8.5) [] {};

\node[fill=red, square, draw=black, inner sep=0pt, minimum size=0.2cm] (a1s1) at (-6,3) [label=right:${a_2^{\square s_1}}$] {}; 
%\node[fill=red, square, draw=black] (b1ss1) at (-7,4) [label=above:$b_1^{\square s_1}$] {}; 
%\vertex (a2^s1) at (-6.2, 4) [label=above:$a_2^{s_1}$] {}; 
%\vertex (b2^s1) at (-5.5, 4) [label=above:$b_2^{s_1}$] {}; 

\node[fill=red, square, draw=black, inner sep=0pt, minimum size=0.2cm] (a0s1) at (-8.5,3) [label=left:${a_1^{\square s_1}}$] {}; 
\node[fill=red, square, draw=black, inner sep=0pt, minimum size=0.2cm] (b0ss1) at (-9.3,4) [label=above:$b_1^{\square s_1}$] {}; 
\vertex (a0^s1) at (-8.5, 4) [label=above:$a_1^{s_1}$] {}; 
\vertex (b0^s1) at (-7.9, 4) [label=above:$b_1^{s_1}$] {};

\node[fill=red, square, draw=black, inner sep=0pt, minimum size=0.2cm] (b2ss1) at (-4.6,4) [label=above:$b_3^{\square s_1}$] {}; 
\vertex (a2^s1) at (-3.9, 4) [label=above:$a_3^{s_1}$] {}; 
\vertex (b2^s1) at (-3.1, 4) [label=above:$b_3^{s_1}$] {}; 

\node[fill=red, square, draw=black, inner sep=0pt, minimum size=0.2cm] (a2s1) at (-4,3) [label=right:${a_3^{\square s_1}}$] {};   
\node[fill=red, square, draw=black, inner sep=0pt, minimum size=0.2cm] (b1ss1) at (-7,4) [label=above:$b_2^{\square s_1}$] {}; 
\vertex (a1^s1) at (-6.2, 4) [label=above:$a_2^{s_1}$] {}; 
\vertex (b1^s1) at (-5.5, 4) [label=above:$b_2^{s_1}$] {}; 

\vertex (xs1) at (-5, 2) [label=left:$x_{s_1}$] {}; 
\node[fill=red, square, draw=black, inner sep=0pt, minimum size=0.2cm] (zs1) at (-5,0) [label=left:${z_{s_1}}$] {}; 
\vertex (ys1) at (-5, -2) [label=left:$y_{s_1}$] {};
\node[fill=red, square, draw=black, inner sep=0pt, minimum size=0.2cm] (s1z0) at (-4,1) {};  
\node[triangle, draw=black, fill= blue, inner sep=0pt, minimum size=0.2cm] (s1z1) at (-4, .5) [] {}; 
\node[triangle, draw=black, fill= blue, inner sep=0pt, minimum size=0.2cm] (s1z2) at (-4, 0) [] {}; 
\node[triangle, draw=black, fill= blue, inner sep=0pt, minimum size=0.2cm] (s1z3) at (-4, -0.5) [] {}; 
\node[triangle, draw=black, fill= blue, inner sep=0pt, minimum size=0.2cm] (s1z4) at (-4, -1) [] {}; 
\node[triangle, draw=black, fill= blue, inner sep=0pt, minimum size=0.2cm] (s1z5) at (-4, -1.5) [] {}; 

\vertex (c1s1) at (-6.5, -3) [label=below:$c_1^{s_1}$] {};
\vertex (c2s1) at (-5.5, -3) [label=below:$c_2^{s_1}$] {};
\vertex (c3s1) at (-4.5, -3) [label=below:$c_3^{s_1}$] {};
\vertex (c4s1) at (-3.5, -3) [label=below:$c_4^{s_1}$] {};

\node[fill=red, square, draw=black, inner sep=0pt, minimum size=0.2cm] (a) at (0,-6) [label= left:${a}$] {};

\node[triangle, draw=black, fill= blue, inner sep=0pt, minimum size=0.2cm] (a0) at (-1.2,-7) [label=below:${a_1^{\triangle}}$] {};
\node[triangle, draw=black, fill= blue, inner sep=0pt, minimum size=0.2cm] (a1) at (-0.5,-7) [label=below:${a_2^{\triangle}}$] {};
\node[triangle, draw=black, fill= blue, inner sep=0pt, minimum size=0.2cm] (a2) at (0.2,-7) [label=below:${a_3^{\triangle}}$] {};
\node[fill=red, square, draw=black, inner sep=0pt, minimum size=0.2cm] (a3) at (1,-7) [label=below:${a^{\square}}$] {};

\path
(a) edge (a1)
(a) edge (a2)
(a) edge (a3)
(a) edge (a0)
(xs1) edge (a0s1)
(xs1) edge (zs1)
(ys1) edge (zs1)
(xs1) edge (a1s1)
(xs1) edge (a2s1)
(zs1) edge (s1z0)
(zs1) edge (s1z1)
(zs1) edge (s1z2)
(zs1) edge (s1z3)
(zs1) edge (s1z4)
(zs1) edge (s1z5)
(ys1) edge (c1s1)
(ys1) edge (c2s1)
(ys1) edge (c3s1)
(ys1) edge (c4s1)
(a1s1) edge (b1ss1)
(a1s1) edge (b1^s1)
(a1s1) edge (a1^s1)
(a2s1) edge (b2ss1)
(a2s1) edge (b2^s1)
(a2s1) edge (a2^s1)
(u1) edge (u11)
(u1) edge (u12)
(u1) edge (u13)
(u1) edge (u14)
(u2) edge (u21)
(u2) edge (u22)
(u2) edge (u23)
(u2) edge (u24)
(u2) edge (ut1)
(u2) edge (ut2)
(u2) edge (ut3)
(u2) edge (ut4)
(u1) edge (ut11)
(u1) edge (ut12)
(u1) edge (ut13)
(u1) edge (ut14)
(a0s1) edge (b0ss1)
(a0s1) edge (a0^s1)
(a0s1) edge (b0^s1);

\node[fill=red, square, draw=black, inner sep=0pt, minimum size=0.2cm] (a1s2) at (-1,3) [label=right:${a_1^{\square s_2}}$] {};

\node[fill=red, square, draw=black, inner sep=0pt, minimum size=0.2cm] (a2s2) at (1,3) [label=right:${a_2^{\square s_2}}$] {}; 

\node[fill=red, square, draw=black, inner sep=0pt, minimum size=0.2cm] (b1ss2) at (-2,4) [label=above:$b_1^{\square s_2}$] {}; 
\vertex (a1^s2) at (-1.2, 4) [label=above:$a_1^{s_2}$] {}; 
\vertex (b1^s2) at (-0.5, 4) [label=above:$b_1^{s_2}$] {}; 

\node[fill=red, square, draw=black, inner sep=0pt, minimum size=0.2cm] (b2ss2) at (0.4,4) [label=above:$b_2^{\square s_2}$] {}; 
\vertex (a2^s2) at (1.1, 4) [label=above:$a_2^{s_2}$] {}; 
\vertex (b2^s2) at (1.9, 4) [label=above:$b_2^{s_2}$] {};

\node[fill=red, square, draw=black, inner sep=0pt, minimum size=0.2cm] (zs2) at (0,0) [label=left:${z_{s_2}}$] {}; 
\vertex (xs2) at (0, 2) [label=left:$x_{s_2}$] {};
\vertex (ys2) at (0, -2) [label=left:$y_{s_2}$] {};

\node[fill=red, square, draw=black, inner sep=0pt, minimum size=0.2cm] (s2z0) at (1,1) {};  
\node[triangle, draw=black, fill= blue, inner sep=0pt, minimum size=0.2cm] (s2z1) at (1, .5) [] {}; 
\node[triangle, draw=black, fill= blue, inner sep=0pt, minimum size=0.2cm] (s2z2) at (1, 0) [] {}; 
\node[triangle, draw=black, fill= blue, inner sep=0pt, minimum size=0.2cm] (s2z3) at (1, -0.5) [] {}; 
\node[triangle, draw=black, fill= blue, inner sep=0pt, minimum size=0.2cm] (s2z4) at (1, -1) [] {}; 
\node[triangle, draw=black, fill= blue, inner sep=0pt, minimum size=0.2cm] (s2z5) at (1, -1.5) [] {}; 
\vertex (c1s2) at (-1.5, -3) [label=below:$c_1^{s_2}$] {};
\vertex (c2s2) at (-0.5, -3) [label=below:$c_2^{s_2}$] {};
\vertex (c3s2) at (0.5, -3) [label=below:$c_3^{s_2}$] {};
\vertex (c4s2) at (1.5, -3) [label=below:$c_4^{s_2}$] {};

\path
(xs2) edge (zs2)
(ys2) edge (zs2)
(xs2) edge (a1s2)
(xs2) edge (a2s2)
(zs2) edge (s2z0)
(zs2) edge (s2z1)
(zs2) edge (s2z2)
(zs2) edge (s2z3)
(zs2) edge (s2z4)
(zs2) edge (s2z5)
(ys2) edge (c1s2)
(ys2) edge (c2s2)
(ys2) edge (c3s2)
(ys2) edge (c4s2)
(a1s2) edge (b1ss2)
(a1s2) edge (b1^s2)
(a1s2) edge (a1^s2)
(a2s2) edge (b2ss2)
(a2s2) edge (b2^s2)
(a2s2) edge (a2^s2);

\node[fill=red, square, draw=black, inner sep=0pt, minimum size=0.2cm] (a1s3) at (3.5,3) [label=right:${a_1^{\square s_3}}$] {}; 
\node[fill=red, square, draw=black, inner sep=0pt, minimum size=0.2cm] (a2s3) at (5.5,3) [label=right:${a_2^{\square s_3}}$] {}; 
\node[fill=red, square, draw=black, inner sep=0pt, minimum size=0.2cm] (a3s3) at (7.5,3) [label=right:${a_3^{\square s_3}}$] {}; 

\node[fill=red, square, draw=black, inner sep=0pt, minimum size=0.2cm] (b1ss3) at (2.7,4) [label=above:$b_1^{\square s_3}$] {}; 
\vertex (a1^s3) at (3.4, 4) [label=above:$a_1^{s_3}$] {}; 
\vertex (b1^s3) at (4, 4) [label=above:$b_1^{s_3}$] {}; 

\node[fill=red, square, draw=black, inner sep=0pt, minimum size=0.2cm] (b2ss3) at (4.8,4) [label=above:$b_2^{\square s_3}$] {}; 
\vertex (a2^s3) at (5.5, 4) [label=above:$a_2^{s_3}$] {}; 
\vertex (b2^s3) at (6.2, 4) [label=above:$b_2^{s_3}$] {}; 

\node[fill=red, square, draw=black, inner sep=0pt, minimum size=0.2cm] (b3ss3) at (7,4) [label=above:$b_3^{\square s_3}$] {}; 
\vertex (a3^s3) at (7.7, 4) [label=above:$a_3^{s3}$] {}; 
\vertex (b3^s3) at (8.4, 4) [label=above:$b_3^{s_3}$] {}; 

\node[fill=red, square, draw=black, inner sep=0pt, minimum size=0.2cm] (zs3) at (5,0) [label=left:${z_{s_3}}$] {}; 
\vertex (xs3) at (5, 2) [label=left:$x_{s_3}$] {};
\vertex (ys3) at (5, -2) [label=left:$y_{s_3}$] {};
\node[fill=red, square, draw=black, inner sep=0pt, minimum size=0.2cm] (s3z0) at (6,1) {};  
\node[triangle, draw=black, fill= blue, inner sep=0pt, minimum size=0.2cm] (s3z1) at (6, .5) [] {}; 
\node[triangle, draw=black, fill= blue, inner sep=0pt, minimum size=0.2cm] (s3z2) at (6, 0) [] {}; 
\node[triangle, draw=black, fill= blue, inner sep=0pt, minimum size=0.2cm](s3z3) at (6, -0.5) [] {}; 
\node[triangle, draw=black, fill= blue, inner sep=0pt, minimum size=0.2cm] (s3z4) at (6, -1) [] {}; 
\node[triangle, draw=black, fill= blue, inner sep=0pt, minimum size=0.2cm] (s3z5) at (6, -1.5) [] {}; 
\vertex (c1s3) at (2.5, -3) [label=below:$c_1^{s_3}$] {};
\vertex (c2s3) at (3.5, -3) [label=below:$c_2^{s_3}$] {};
\vertex (c3s3) at (4.5, -3) [label=below:$c_3^{s_3}$] {};
\vertex (c4s3) at (5.5, -3) [label=below:$c_4^{s_3}$] {};
\vertex (c5s3) at (6.5, -3) [label=below:$c_5^{s_3}$] {};
\vertex (c6s3) at (7.5, -3) [label=below:$c_6^{s_3}$] {};
\path
(xs3) edge (zs3)
(ys3) edge (zs3)
(xs3) edge (a1s3)
(xs3) edge (a2s3)
(xs3) edge (a3s3)
(zs3) edge (s3z0)
(zs3) edge (s3z1)
(zs3) edge (s3z2)
(zs3) edge (s3z3)
(zs3) edge (s3z4)
(zs3) edge (s3z5)
(ys3) edge (c1s3)
(ys3) edge (c2s3)
(ys3) edge (c3s3)
(ys3) edge (c4s3)
(ys3) edge (c5s3)
(ys3) edge (c6s3)
(a1s3) edge (b1ss3)
(a1s3) edge (b1^s3)
(a1s3) edge (a1^s3)
(a2s3) edge (b2ss3)
(a2s3) edge (b2^s3)
(a2s3) edge (a2^s3)
(a3s3) edge (b3ss3)
(a3s3) edge (b3^s3)
(a3s3) edge (a3^s3)
(u1) edge (a1^s1)
(u1) edge (a0^s1)
(u1) edge (a1^s2)
(u1) edge (a1^s3)
(u2) edge (a1^s1)
(u2) edge (a2^s2)
(u2) edge (a1^s3)
(u2) edge (a2^s3)
(a) edge (c1s1)
(a) edge (c2s1)
(a) edge (c3s1)
(a) edge (c4s1)
(a) edge (c1s2)
(a) edge (c2s2)
(a) edge (c3s2)
(a) edge (c4s2)
(a) edge (c1s3)
(a) edge (c2s3)
(a) edge (c3s3)
(a) edge (c4s3)
(a) edge (c5s3)
(a) edge (c6s3);

\path 
(a0^s1) [green] edge (a)
(b0^s1) [green] edge (a)
(a1^s1) [green] edge (a)
(b1^s1) [green] edge (a)
(a2^s1) [green] edge (a)
(b2^s1) [green] edge (a)
(a2^s2) [green] edge (a)
(b2^s2) [green] edge (a)
(a1^s2) [green] edge (a)
(b1^s2) [green] edge (a)
(a1^s3) [green] edge (a)
(b1^s3) [green] edge (a)
(a2^s3) [green] edge (a)
(b2^s3) [green] edge (a)
(a3^s3) [green] edge (a)
(b3^s3) [green] edge (a);

 \end{tikzpicture}
 \caption{The graph $G$ in the proof of Lemma \ref{oatheorem1} constructed for MRSS instance 
 $S=\{(2, 1), (1,1), (1,2)\}$, $t=(3,3)$, $k=2$ and $k'=2$.}
     \label{oafig1}
\end{figure}

 \proof To prove this we reduce from MRSS, which  is known to be W[1]-hard when 
parameterized by the combined parameter $k+k'$, even if all integers in the input are given in unary \cite{mss}.
Let  $I = (k, k', S, t)$  be an instance of {\sc MRSS}. We construct an  instance 
 $I'=(G,r,V_{\triangle},V_{\square})$ of {\sc Strong Offensive Alliance$^{\mbox{FN}}$} in the following way.
 See Figure \ref{oafig1} for an illustration.
 \noindent First, we  introduce a  set  of $k$ forbidden vertices $U=\{ u_1,u_2,\ldots,u_k\}$. For each vector
$s=(s(1),s(2),\ldots,s(k))\in S$, we introduce a tree $T_{s}$ into $G$.  We define $\mbox{max}(s)=\max\limits_{1\leq i \leq k}\{s(i)\}$.
The vertex set of tree $T_s$ is defined as follows:
$$ V(T_s) = A_s \cup B_s \cup A^{\square}_s \cup B^{\square}_s \cup C_{s} \cup Z_{s}  
\cup \Big\{x_s,y_s,z_s\Big\} $$
where  $A_s=\{a^{s}_{1},\ldots,a^s_{\max(s)+1}\}$, 
$B_{s}=\{b^{ s}_{1},\ldots,b^{ s}_{\max(s)+1}\}$, 
$ A_{s}^{\square}=\{a^{\square s}_{1},\ldots,a^{\square s }_{\max(s)+1}\}$, 
$B_s^{\square}=\{b^{ \square s}_{1},\ldots,b^{\square s }_{\max(s)+1}\}$ and 
$C_{s}=\{c_{1}^{s},\ldots,c_{2\max{(s)+2}}^{s}\}$ are five sets of vertices, and 
%such that 
%$|A_s|=|B_{s}|=|A_{s}^{\square}|=|B_{s}^{\square}|=\mbox{max}(s)+1$ 
%and $|C_{s}|=2\mbox{max}(s)+2$; 
the set $Z_{s}=\{z^{\triangle s}_{1},z^{\triangle s}_{2},z^{\triangle s}_{3},
z^{\triangle s}_{4},z^{\triangle s}_{5},
z^{\square s} \}$ contains five necessary vertices and one forbidden vertex. 
We now create the edge set of $T_s$.
\begin{align*}
    E(T_s) &=  \bigcup\limits_{i=1}^{{\max(s)+1}} 
    \Big\{ (a_{i}^{\square s},b_{i}^{\square s}), (a_{i}^{\square s},a_{i}^{s}),
    (a_{i}^{\square s},b_{i}^{s}) ,(x_s,a_{i}^{\square s})\Big\} \\ 
    & \bigcup\limits_{i=1}^{5} \{(z_s,z_i^{\triangle s}),(z_s,z^{\square s})\} 
    \bigcup \{(x_s,z_s),(z_s,y_s)\} 
     \bigcup\limits_{i=1}^{2\max(s)+2} (y_s,c_i^s)
\end{align*}

%\begin{align*}
%    E(T_s) &= \bigcup\limits_{i=1}^{n} \bigcup\limits_{j=1}^{{\max(s_{i})+1}} 
%    \{ (l_{s_{i}}^{\square j},l_{s_{i}}^{\prime \square j}), (l_{s_{i}}^{\square j},l_{s_{i}}^{ j}), (l_{s_{i}}^{\square j},l_{s_{i}}^{\prime j}),(x_{s_{i}},l_{s_{i}}^{\sqaure j}) \} \\ & \bigcup\limits_{i=1}^{n}\bigcup\limits_{q=1}^{5} \{(z_{s_{i}},z_{s_{i}}^{q \triangle}),(z_{s_{i}},z^{\square}_{s_{i}})\} \bigcup \{(x_{s_{i}},z_{s_{i}}),(x_{s_{i}},y_{s_{i}})\} 
 %   \\ & \bigcup\limits_{i=1}^{n}\bigcup\limits_{t=1}^{2.\max(s_{i})+2} (y_{s_{i}},p^{t}_{s_{i}})
%\end{align*}
\noindent  Next we introduce a vertex $a$ and a set of four vertices $A=\{a_{1}^{\triangle},a_{2}^{\triangle},a_{3}^{\triangle},
 a^{\square}\}$ containing three necessary vertices and one forbidden vertex. Make $a$ adjacent to 
 all the vertices in $A$.
  For each $i\in \{1,2,\ldots,k\}$ and 
 for each $s\in S$, we make $u_{i}$ adjacent to exactly $s(i)$ many vertices 
 of  $A_s$ in arbitrary manner.  For each $s\in S$, we make $a$ adjacent to all  
 the vertices of  $A_s\cup B_{s} \cup  C_s$. 
 For every $u_i\in U$, we create a set $V_{u_i\square}$ of $\sum\limits_{s\in S}{s(i)}$ 
 forbidden vertices  and a set $V_{u_i\triangle}$ of $2\sum\limits_{s\in S}{s(i)}-2t(i)+2$ 
 necessary vertices; and make $u_i$ adjacent to every vertex 
 of $V_{u_i\square}\cup V_{u_i\triangle}$.
We define $$V_{\triangle} = \bigcup\limits_{i=1}^{k} V_{u_i\triangle} \bigcup A \setminus \{a^{\square}\} 
\bigcup\limits_{s\in S} {Z_{s} \setminus \{ z^{\square s}\} }   $$ and 
$$V_{\square} = U\cup \{a,a^{\square}\} \bigcup\limits_{i=1}^{k} V_{u_i\square} \bigcup\limits_{s\in S} 
A_s^{\square}\cup B_s^{\square}\cup \{z_s, z^{\square s}\}.$$
We set $r= \sum\limits_{i=1}^{k} {2\Big(\sum\limits_{s\in S}{s(i)}-t(i)+1\Big)}+\sum\limits_{s\in S}2(\max(s)+1)+5n+3+k'$.
Observe that if we remove the set $U\cup \{a\}$ of $k+1$ vertices from $G$, each 
connected component 
of the resulting graph is a tree with height at most 5. Note that, $I'$ can be constructed in polynomial time. The reason is this. As 
all integers in $I$ are bounded by a polynomial in $n$, the number of vertices in $G$ is also polynomially  bounded in $n$.

\par It remains to show that  $I$ is a yes instance if and only if $I'$ is a yes instance. 
Towards showing the forward direction, let $S'$ be a subset of $S$ such that $|S'|\leq k'$ and $\sum\limits_{s\in S'}{s}\geq t$. 
We claim 
$$ R = V_{\triangle} \bigcup\limits_{s\in S'} A_{s}\cup B_{s}\cup \{x_{s}\} \bigcup\limits_{s\in S\setminus S'} C_{s} $$
is a strong offensive alliance of $G$ such that $|R|\leq r$, $V_{\triangle}\subseteq R$, and $V_{\square} \cap R=\emptyset$. 
Observe that 
$N_{G}(R) = U \cup \{a\} \bigcup\limits_{s\in S} \{z_s\} \bigcup\limits_{s\in S\setminus S'} \{y_{s}\} \bigcup\limits_{s\in S'} A_{s}^{\square} $. 
%We  show that for each vertex $x\in N_{G}(R)$, $d_{R}(x) \geq d_{R^c}(x)+1$. 
Let $u_i \in U$, 
then we show that $d_{R}(u_{i})\geq d_{R^c}(u_{i})+2$. 
As $\sum\limits_{s \in S'} s(i)-t(i)\geq 0$, we get
\begin{equation*}
\begin{split}
 d_{R}(u_i)&=\sum\limits_{s \in S'} s(i) + |V_{u_i\triangle}|\\
           &=\sum\limits_{s \in S'} s(i)+2\sum\limits_{s\in S}{s(i)}- 2t(i)+2\\
           &=\Big(\sum\limits_{s \in S'} s(i)-t(i)\Big )+ \sum\limits_{s\in S}{s(i)}-t(i)+ \sum\limits_{s\in S}{s(i)}+2 \\
           &\geq \sum\limits_{s\in S}{s(i)} -t(i) + \sum\limits_{s\in S}{s(i)}+2 \\
           &=\sum\limits_{s\in S \setminus S'}{s(i)} +\Big(\sum\limits_{s\in S'}{s(i)} -t(i) \Big)+ \sum\limits_{s\in S}{s(i)} +2\\
           &\geq  \sum\limits_{s\in S \setminus S'}{s(i)}+\sum\limits_{s\in S}{s(i)}+2
           =\sum\limits_{s\in S \setminus S'}{s(i)}+|V_{u_i\square}|+2\\
           &=d_{R^c}(u_i)+2.
 \end{split}          
\end{equation*}
For the remaining vertices $x$ in $N(R)$, 
it is easy to see that $d_{R}(x) \geq d_{R^c}(x)+2$. 
Therefore, $R$ is a strong offensive alliance.

\par Towards showing the reverse direction of the equivalence, suppose 
$G$ has a strong offensive alliance 
$R$ of size at most $r$  such that $V_{\triangle} \subseteq R$ 
and $V_{\square}\cap R=\emptyset$. From the definition of $V_{\triangle}$
and $V_{\square}$, it is easy to note that $U \subseteq N(R)$. 
We know $V_{\triangle}$ contains $\sum\limits_{i=1}^{k} {\Big(\sum\limits_{s\in S}{2s(i)}-2t(i)\Big)+5n+3}$ vertices; thus
besides the vertices of $V_{\triangle}$, there are 
at most $\sum\limits_{s\in S}2(\max(s)+1)+k'$  vertices in $R$.
Since $a\in N(V_{\triangle})$ and $d_{G}(a) = \sum\limits_{s\in S}4(\max(s)+1) +4$ where $a$ is adjacent to three necessary vertices, it must have at least $\sum\limits_{s\in S}2(\max(s)+1)$ many neighbours 
in $R$ from the set $\bigcup\limits_{s\in S} (A_{s}\cup B_{s}\cup C_s)$. 
It is to be noted that if a vertex from the set $A_{s}\cup B_s$ 
is in the solution  then the whole set $A_{s}\cup B_s\cup \{x_s\}$ lie in the solution. 
Otherwise  $v\in A_{s}^{\square}\subseteq N(R)$ will have $d_R(v)<d_{R^c}(v)+2$ 
which is a contradiction as $R$ is a strong offensive alliance. 
This shows that at most $k'$ many sets of the form $A_s\cup B_s\cup \{x_s\}$  contribute to the solution as 
otherwise the size of solution  exceeds $r$. 
Therefore, any  strong offensive alliance $R$ of size at most 
$r$  can be transformed to another strong offensive alliance $R'$ of size at most $r$ as follows:
$$ R' =  V_{\triangle} \bigcup\limits_{x_{s}\in R} A_{s}\cup B_s\cup \{x_{s}\} \bigcup\limits_{x_{s}\in V(G)\setminus R} C_{s}.$$ 
%Thus  $N_{G}(R) = U\cup \{b\} \bigcup\limits_{x^{s}\in R'} \{y^{s}\}\cup X^{s}_{\square} \bigcup\limits_{x^{s}\notin R'} \{z^{s}\}$. %Clearly for $ u \in U$, we have $ d_{R}(u)\geq d_{R'}(u) \geq d_{R'^c}(u)+1 \geq d_{R^c}(u)+1$. 
%For the rest of the vertices  $v\in N_{G}(R)$, we can easily see that $d_{R}(v) \geq d_{R^c}(v)+1$. 
We define a subset $S'=\Big\{ s \in S ~|~ x_s\in R'\Big\}.$
Clearly, $|S'|\leq k'$. We claim that $\sum\limits_{s\in S'} s(i) \geq t(i)$ for all $1 \leq i \leq k$. 
 Assume for the sake of contradiction that $\sum\limits_{s\in S'} s(i) < t(i)$ for some $i \in \{1,2,\ldots,k\}$. 
Then, we have 
\begin{equation*}
\begin{split}
 d_{R'}(u_i)&=\sum\limits_{s \in S'} {s(i)} +|V_{u_i\triangle}|\\
           &=\sum\limits_{s \in S'} {s(i)}+2\sum\limits_{s\in S}{s(i)}- 2t(i)+2\\
           &=\sum\limits_{s \in S'} {s(i)}-t(i) +\sum\limits_{s\in S}{s(i)}- t(i) +\sum\limits_{s\in S}{s(i)}+2\\
           &< \sum\limits_{s\in S}{s(i)} -t(i) +\sum\limits_{s\in S}{s(i)} +2\\
           &= \sum\limits_{s\in S\setminus S'}{s(i)}+\Big(\sum\limits_{s\in S'}{s(i)} -t(i)\Big) +\sum\limits_{s\in S}{s(i)}+2\\
           &<  \sum\limits_{s\in S \setminus S'}{s(i)} +\sum\limits_{s\in S}{s(i)}+2=\sum\limits_{s\in S \setminus S'}{s(i)} +|V_{u_i\square}|+2\\
           &=d_{R'^c}(u_i)+2
 \end{split}          
\end{equation*}
and we also know  $u_{i}\in N(R')$, which is a contradiction to the fact that $R'$ is a strong offensive alliance. This shows that $I$ is a yes instance. \qed \\

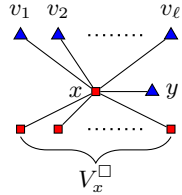
\begin{figure}[H]
\centering
\[\begin{tikzpicture}[scale=0.5][H]

\centering

\node [fill=red, square, draw=black, inner sep=0pt, minimum size=0.15cm] (a) at (0, .5) [label=left:$x$]{};

\node [triangle, draw=black, fill= blue, inner sep=0pt, minimum size=0.2cm] (y) at (1.5, 0.5) [label=right:$y$]{};

\node[fill=red, square, draw=black, inner sep=0pt, minimum size=0.15cm](a1) at (-2, -0.5) []{};
\node[fill=red, square, draw=black, inner sep=0pt, minimum size=0.15cm](b7) at (-1, -0.5) []{};
\node[fill=red, square, draw=black, inner sep=0pt, minimum size=0.15cm] (a2) at (2, -0.5) []{};
\node [triangle, draw=black, fill= blue, inner sep=0pt, minimum size=0.2cm] (b1) at (-2, 2) [label=above:$v_{1}$]{};
\node [triangle, draw=black, fill= blue, inner sep=0pt, minimum size=0.2cm] (b5) at (-1, 2) [label=above:$v_{2}$]{};
%\node [fill=red, square, draw] (b6) at (-1, 3) [label=above:$v_{2}^{\square}$]{};
%\node [fill=red, square, draw] (b3) at (-2, 3) [label=above:$v_{1}^{\square}$]{};
\node [triangle, draw=black, fill= blue, inner sep=0pt, minimum size=0.2cm] (b2) at (2, 2) [label=above:$v_{\ell}$]{};
%\node [fill=red, square, draw] (b4) at (2, 3) [label=above:$v_{l}^{\square}$]{};

\node (t) at (0, -1) [label=below:$V_{x}^{\square}$]{};

\path 
(a) edge (y)
%(b1) edge (b3)
%(b6) edge (b5)
%(b2) edge (b4)
(a) edge (b7)
(a) edge (b5)
(a) edge (a1)
(a) edge (a2)
(a) edge (b1)
(a) edge (b2);

\draw[dotted, thick] (-0.2,-0.5) -- (1.3,-0.5);
\draw[dotted, thick] (-.2,2) -- (1.3,2);

\draw [decorate,decoration={brace,amplitude=10pt,mirror},xshift=0pt,yshift=0pt] (-2,-0.7) --  (2,-0.7) node [black,midway,xshift=-0.6cm]{};

\end{tikzpicture}
\]
\caption{An illustration of the gadget used in the proof of Corollary \ref{corollary}.}
\label{CorollaryFig1}
\end{figure}

\noindent We have the following corollaries  from Lemma \ref{oatheorem1}.
 \begin{corollary}\label{corollary}\rm
 The {\sc Strong Offensive Alliance$^{\mbox{FN}}$} problem is W[1]-hard when parameterized by  the size of 
 a vertex deletion set into trees of height at most 5, even when $|V_{\triangle}|=1$. 
 \end{corollary}
  \proof Given an instance $I=(G,r,V_{\triangle},V_{\square})$ of {\sc Strong  Offensive Alliance$^{\mbox{FN}}$},
 we construct an equivalent instance $I'=(G',r',V_{\triangle}^{\prime},V_{\square}^{\prime})$ 
 with $|V_{\triangle}^{\prime}|=1$. See Figure \ref{CorollaryFig1} for an illustration. 
\noindent Let $v_1,v_2,\dots,v_{\ell}$ be vertices of $V_{\triangle}$ where we assume that $\ell>1$.
 We introduce two vertices $x$ and $y$ where $x$ is a forbidden vertex and $y$ is a necessary vertex;
 and make $x$ and $y$ adjacent. 
 We make $x$ adjacent to all the vertices in $V_{\triangle}$. 
 We also introduce a set $V_{x\square}$ of $\ell-1$  forbidden vertices and make them  adjacent to $x$. 
Set $r'=r+1$.  Define $V_{\triangle}^{\prime}=\{y\}$ and $V_{\square}^{\prime}=\{x \}\cup V_{x\square} \cup V_{\square}$. 
 We also define $G'$ as follows $$ V(G') = V(G) \cup \{x,y\} \cup V_{x\square} $$ and $$ E(G') = E(G) \bigcup \Big\{(x,y),(x,\alpha),(x,\beta) ~|~ \alpha\in V_{x\square},\beta \in V_{\triangle} \Big\}.$$  
 Let $H$  be a vertex deletion set   of $G$ into trees of height at most 5. 
 Clearly, if $H$ has at most $k$ vertices then the set $H\cup \{x\}$ has at most $k+1$ vertices and  is a vertex deletion set
 of $G'$ into trees of height at most 5. 
 It is easy to see that $I$ and $I'$ are equivalent  instances.\qed\\\\
 
 \noindent We can get an analogous result for the exact variant.
 \begin{corollary}\label{corollaryExactSOAFN}\rm
 The {\sc Exact Strong Offensive Alliance$^{\mbox{FN}}$} problem is W[1]-hard when parameterized by  the size of a vertex deletion set into trees of height at most 5 even when $|V_{\triangle}|=1$.
 \end{corollary}
 
 Next, we give an FPT reduction that eliminates necessary vertices.
 
\begin{lemma}\label{twtheorem3}\rm
 The {\sc Offensive Alliance$^{\mbox{F}}$} problem is W[1]-hard when parameterized by the size of a vertex deletion set into trees of height at most 5.
 \end{lemma}
 
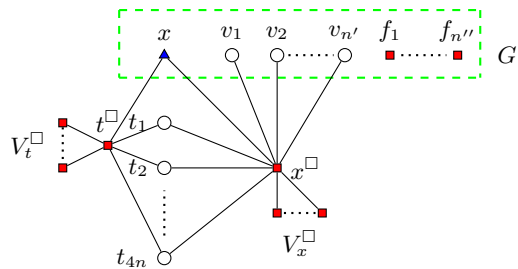
\begin{figure}[ht]
\centering
 \begin{tikzpicture}[scale=0.3]
 
\draw[green, dashed, thick] (-7,7) -- (9,7);
\draw[green, dashed, thick] (-7,7) -- (-7,4);
\draw[green, dashed, thick] (-7,4) -- (9,4);
\draw[green, dashed, thick] (9,7) -- (9,4);
 
\node[fill=red, square, draw=black, inner sep=0pt, minimum size=0.15cm] (x00) at (0,0) [label= right:${x^{\square}}$] {};

\vertex (v1) at (-2,5) [label=above:${v_{1}}$] {};
\vertex (v2) at (0,5) [label=above:${v_{2}}$] {};
\vertex (v3) at (3,5) [label=above:${v_{n'}}$] {};

\node[fill=red, square, draw=black, inner sep=0pt, minimum size=0.15cm] (f1) at (5,5) [label=above:${f_{1}}$] {};
\node[fill=red, square, draw=black, inner sep=0pt, minimum size=0.15cm] (f2) at (8,5) [label=above:${f_{n''}}$] {};

\draw[dotted, thick] (5.5,5) -- (7.5,5);
\draw[dotted, thick] (0.5,5) -- (2.5,5);

\draw[dotted, thick] (-5,-3) -- (-5,-1);

\node (g) at (9,5) [label=right:${G}$] {};

\node[fill=red, square, draw=black, inner sep=0pt, minimum size=0.15cm] (x01) at (0,-2) [] {};

\draw[dotted, thick] (0.3,-2) -- (1.7,-2);

\node[fill=red, square, draw=black, inner sep=0pt, minimum size=0.15cm] (x02) at (2,-2) [] {};

\node (vx) at (1,-2) [label= below:$V_{x}^{\square}$] {};

\node [triangle, draw=black, fill= blue, inner sep=0pt, minimum size=0.15cm] (x) at (-5,5) [label= above:${x}$] {};

\vertex (y1) at (-5,2) [label= left:${t_{1}}$] {};
\vertex (y2) at (-5,0) [label= left:${t_{2}}$] {};
\vertex (yn) at (-5,-4) [label= left:${t_{4n}}$] {};

\node[fill=red, square, draw=black, inner sep=0pt, minimum size=0.15cm] (t) at (-7.5,1) [label=above:${t^{\square}}$] {};

\node[fill=red, square, draw=black, inner sep=0pt, minimum size=0.15cm] (t0) at (-9.5,2) [] {};
\node[fill=red, square, draw=black, inner sep=0pt, minimum size=0.15cm] (t1) at (-9.5,0) [] {};

\node (T) at (-9.5,1) [label=left:${V_{t}^{\square}}$] {};

\draw[dotted, thick] (-9.5,1.8) -- (-9.5,0.2);

\path 

(x00) edge (x01)
(x00) edge (x02)
(x00) edge (x)
(x00) edge (v1)
(x00) edge (v2)
(x00) edge (v3)
(x00) edge (y1)
(x00) edge (y2)
(x00) edge (yn)
(t) edge (x)
(t) edge (y1)
(t) edge (y2)
(t) edge (yn)
(t) edge (t0)
(t) edge (t1);

\end{tikzpicture}
    \caption{The reduction from {\sc Strong Offensive Alliance$^{\mbox{FN}}$} to 
    {\sc Offensive Alliance$^{\mbox{F}}$} in Lemma \ref{twtheorem3}. Note that the set $\{v_{1},\ldots,v_{n'}\}$ may contain forbidden vertices of degree greater  than one.}
    \label{oafig2}
\end{figure}

 \proof  To prove this we reduce from the {\sc Strong Offensive Alliance$^{\mbox{FN}}$} problem, which is 
 W[1]-hard when parameterized by  the size of a vertex deletion set into trees of height at most 5, even when $|V_{\triangle}|=1$.
 See Corollary \ref{corollary}. 
Given an instance $I=(G,r,V_{\triangle}=\{x\},V_{\square})$ of {\sc Strong Offensive Alliance$^{\mbox{FN}}$}, 
 we construct an instance $I'=(G',r',V^{\prime}_{\square})$ of {\sc Offensive Alliance$^{\mbox{F}}$} 
 the following way. See Figure \ref{oafig2} for an illustration.
 Let $n$ be the number of vertices in $G$ and 
 let $V(G)=\{x, v_1, v_2,\ldots, v_{n'},f_{1},\ldots,f_{n''}\}$ 
 where $F=\{f_{1},\ldots,f_{n''}\}$ is the set of  degree one forbidden vertices in $V(G)$. 
 We introduce two vertices $t^{\square},x^{\square}$ into $G'$. We create a set 
 $V_{t}^{\square}=\{t^{\square}_{1},\ldots,t^{\square}_{4n}\}$
 of $4n$  forbidden vertices into $G'$ and make them adjacent to $t^{\square}$. We  
 introduce a set 
$V_{x}^{\square}$  of $n$  forbidden vertices and make them  adjacent to
  $x^{\square}$. 
 %For the necessary vertex $x\in V_{\triangle}$, we create a  set
 % $V_{x}^{\square}=\{x'_{1},x'_2,\dots,x'_{n}\}$  of forbidden vertices and
 % make $x$ adjacent to all vertices in  $V_{x}^{\square}$.
Finally we create a set $T=\{t_{1},\ldots,t_{4n}\}$ of $4n$ vertices and 
make the vertices in $T$ adjacent to $t^{\square}$ and $x^{\square}$,
and make the vertices in $V(G)\setminus F$ adjacent to $x^{\square}$. 
We also add an edge $(x,t^{\square})$.
 Set $r'=r+4n$.  We define $G'$ as follows:
 $$ V(G') = V(G) \cup T \cup V_{t}^{\square} \cup V_{x}^{\square} \cup \{t^{\square},x^{\square}\}  $$ and 
\begin{align*}
    E(G') =& E(G)  
    \cup  \Big \{(t^{\square},\alpha)~:~\alpha \in T \cup V_t^{\square} \cup\{x\}\Big\} \\
    & \cup  \Big\{(x^{\square},\beta)~:~\beta \in T\cup V_{x}^{\square} \cup V(G)\setminus F \Big\}
\end{align*}

\noindent  We define $V_{\square}^{\prime} = V_{\square} \cup  V_{t}^{\square} \cup V_{x}^{\square} \cup \{t^{\square},x^{\square}\}$. Observe that there exists a set of at most $k+2$ vertices in $G'$  
whose deletion makes the resulting graph a forest containing trees of height at most 5.  
We can find  such a 
set because there exists a vertex deletion set $H$ of $G$ into trees of height at most 5.  
We  just add $\{x^{\square},t^{\square}\}$ to the set $H$, then the resulting set is of size $k+2$ whose deletion makes the resulting graph a forest containing trees of height at most 5.

 \par We  now claim  that $I$ is a yes-instance if and only if $I'$ is a yes-instance. 
Assume first  that $R$ is a strong offensive alliance of size at most
$r$ in $G$ such that $\{x\} \subseteq R$ and $V_{\square}\cap R=\emptyset$. 
We claim  $R'=R \cup T$ is an offensive alliance of size at most $r+4n$ in $G'$ such that $V^{\prime}_{\square} \cap R'=\emptyset$. Clearly, $N(R') = \{t^{\square},x^{\square}\} \cup N(R)$. For each $v\in N(R)$, 
we know that $d_{R}(v)\geq d_{R^c}(v)+2$ in $G$. Therefore in graph $G'$, 
we get $d_{R'}(v)\geq d_{R'^c}(v)+1$ for each $v\in N(R)$ due to the vertex $x^{\square}$. 
For   $v\in \{x^{\square},t^{\square}\}$, it is clear that $d_{R'}(v)\geq d_{ R'^c}(v)+1$.
This shows that $I'$ is a yes instance. 

\par To prove the reverse direction of the equivalence, suppose  
$R'$ is an offensive alliance of size at most $r'=r+4n$ in $G'$ such that 
$R'\cap V_{\square}^{\prime}=\emptyset$. We claim that $T\cup \{x\} \subseteq R'$. Since $R'$ is non empty, it must contain 
a vertex from the set $T\cup V(G)\setminus F$. Then $x^{\square}\in N(R')$
and it satisfies the condition $d_{R'}(x^{\square})\geq d_{R'^c}(x^{\square})+1$.
Due to $n$ forbidden vertices in the set $V_{x}^{\square}$, node
$x^{\square}$ must have at least $n+1$ neighbours in $R'$. 
This implies that $R'$ contains at least one vertex from $T$.  
Then $t^{\square}\in N(R')$ and it satisfies the condition 
$d_{R'}(t^{\square})\geq d_{{R'}^c}(t^{\square})+1$. Since $|V_{t}^{\square}|=4n$, the
condition $d_{R'}(t^{\square})\geq d_{R'^c}(t^{\square})+1$ forces the set 
$\{x\}\cup T$ to be 
inside the solution. Consider $R=R' \cap V(G)$. Clearly $|R|\leq r$, $x\in R$, 
$R\cap V_{\square}=\emptyset $ and we 
show that $R$ is a strong offensive alliance in $G$. 
For each $v\in N(R') \cap V(G) =N(R)$, we have $N_{R'}(v)\geq N_{R'^c}(v)+1$ in
$G'$.
Notice that we do not have  $x^{\square}$ in $G$ which is  adjacent to all vertices in $N(R)$.  
Thus for each $v\in N(R)$, we get $N_{R}(v)\geq N_{R^c}(v)+2$  in $G$. Therefore $R$ is a strong 
offensive alliance of size at most $r$ in $G$ such that $x\in R$ and 
$R\cap V_{\square}=\emptyset$. This shows that $I$ is a yes instance. \qed \\\\

\begin{corollary}\label{corollary3}\rm
 The {\sc Exact Offensive Alliance$^{\mbox{F}}$} problem is W[1]-hard when parameterized by the size of a vertex deletion set into trees of height at most 5.
 \end{corollary}

\noindent We are now ready to show our main hardness result for {\sc Offensive Alliance} using a reduction
from {\sc Offensive Alliance$^{\mbox{F}}$}.
%Next we give an FPT reduction that eliminates forbidden vertices.

\begin{theorem}\label{twtheorem}\rm
 The {\sc Offensive Alliance} problem is W[1]-hard when parameterized by the size of a vertex deletion set into trees of height at most 7.
 \end{theorem}
 
\proof We  give a parameterized reduction from   {\sc Offensive Alliance$^{\mbox{F}}$} which is 
 W[1]-hard when parameterized by the size of a vertex deletion set into trees of height at most 5. 
Let $I=(G, r, V_{\square})$ be an instance of {\sc Offensive Alliance$^{\mbox{F}}$}. 
Let $n=|V(G)|$. 
We construct an instance $I'=(G',r')$ of {\sc Offensive Alliance} the following way. We
set $r'=r$. Recall that each degree one
forbidden vertex is adjacent to another forbidden vertex and each  forbidden vertex of degree greater than one is adjacent to a degree one forbidden vertex. Let $u$ be a degree one forbidden vertex in $G$
and $u$ is adjacent to another forbidden vertex $v$.  
For each degree one forbidden vertex $u\in V_{\square}$, 
we introduce a tree 
$T_{u}$ rooted at $u$ of height 2 as shown in 
Figure \ref{fig:OA4}. The forbidden vertex $v$ has additional neighbours from the original 
graph $G$ which are not shown in the figure. We define $G'$ as follows: 
$$ V(G') = V(G) \bigcup\limits_{u\in V_{\square}} \Big \{V(T_{u}) ~|~ \text{where } u  \text{ is a  degree one forbidden vertex in } G \Big\} 
$$
and
\begin{align*}
        E(G') = E(G) \bigcup\limits_{u'\in V_{\square}} {E(T_{u})}. 
\end{align*} 
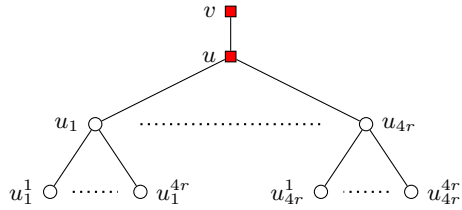
\begin{figure}[ht]
    \centering
 \begin{tikzpicture}[scale=0.6]

\node[fill=red, square, draw=black, inner sep=0pt, minimum size=0.2cm] (h) at (0,0) [label=left:$v$] {};
\node[fill=red, square, draw=black, inner sep=0pt, minimum size=0.2cm] (h1) at (0,-1) [label=left:$u$] {};
\vertex (h11) at (-3,-2.5) [label=left:$u_{1}$] {};
\vertex (h12) at (3,-2.5) [label=right:$u_{4r}$] {};

\vertex (h21) at (-4,-4) [label=left:${u_{1}^{ 1 }}$] {};
\vertex (h22) at (-2,-4) [label=right:${u_{1}^{ 4r }}$] {};

\vertex (h31) at (4,-4) [label=right:${u_{4r}^{ 4r }}$] {};
\vertex (h32) at (2,-4) [label=left:${u_{4r}^{ 1}}$] {};

\path 

(h1) edge (h)
(h1) edge (h11)
(h11) edge (h21)
(h11) edge (h22)
(h1) edge (h12)
(h12) edge (h31)
(h12) edge (h32);

\draw[dotted, thick] (-2,-2.5) -- (2,-2.5);
\draw[dotted, thick] (-3.5,-4) -- (-2.5,-4);
\draw[dotted, thick] (3.5,-4) -- (2.5,-4);

 \end{tikzpicture}
    \caption{Our tree gadget $T_{u}$ for each degree one forbidden vertex $u\in V_{\square}$}
    \label{fig:OA4}
\end{figure}
We claim $I$ is a yes instance if and only if $I'$ is a yes instance. It is easy to see that 
if $R$ is an offensive alliance of size at most $r$ in $G$ such that 
$R\cap V_{\square} =\emptyset$, then it is also an 
offensive alliance of size at most $r'=r$ in $G'$. 
\par To prove the reverse direction of the equivalence, suppose that $G'$ has an  offensive alliance $R'$  of size 
at most $r'=r$. We claim that no vertex from the set $V_{\square} \bigcup\limits_{u\in V_{\square}} V(T_{u})$  
is part of $R'$.
It is easy to see that if any vertex from the set $V_{\square} \bigcup\limits_{u\in V_{\square}} V(T_{u})$ is in $R'$ then the size of $R'$ exceeds $2r$. This implies that $R=R' \cap G$ is an offensive alliance such that $R\cap V_{\square}=\emptyset$ and $|R|\leq r$. This shows that $I$ is a yes instance.\qed\\

We have the following consequences. 
\begin{corollary}\label{corollary4}\rm
 The {\sc Exact Offensive Alliance} problem is W[1]-hard when parameterized by the size of a vertex deletion set into trees of height at most 7.
 \end{corollary}
 
\noindent  Clearly trees of height at most seven are trivially acyclic. 
 Moreover, it is easy to verify that such trees have 
 pathwidth \cite{Kloks94} and treedepth \cite{Sparsity} at most seven, which implies:
 
\begin{theorem}\rm
 The {\sc Offensive Alliance} and {\sc Exact Offensive Alliance} problems
 %and {\sc  offensive alliance$^{\mbox{N}}$} problem even when $|V_{\triangle}|=1$  are 
 are W[1]-hard when parameterized by any of the following parameters:
 \begin{itemize}
     \item the feedback vertex set number,
     \item the treewidth and pathwidth of the input graph,
     \item the treedepth of the input graph.
 \end{itemize}
\end{theorem}

\section{FPT Lower Bound Parameterized by Solution Size}
We know that {\sc Offensive Alliance} admits an FPT algorithm  when parameterized 
by the solution size \cite{Enciso2009AlliancesIG}. The algorithm in \cite{Enciso2009AlliancesIG} uses branching technique and solves the 
problem in $\mathcal{O}^{*}(2^{\mathcal{O}(k \log k)})$ time. It appears that 
this running time is essentially optimal assuming ETH, which is proved 
in the following theorem. Hardness for {\sc Offensive Alliance}  follows from 
a reduction from $k\times k$ {\sc (Permutation) Hitting Set with Thin Sets}. 
In the $k \times k$ {\sc (Permutation) Hitting Set} problem, 
 we are given a family $\mathcal{F}$ of subsets of 
 $[k] \times [k]$, and we would like to find a set $X$, consisting of one vertex 
 from each row and induces a permutation of $[k]$, such that $X \cap F \neq \emptyset$ for each 
 $F\in \mathcal{F}$. In the \emph{thin set}
variant we assume that each $F\in \mathcal{F}$ contains at most one vertex from each row. 
In the proof, we will use the fact that  $k\times k$ {\sc (Permutation) Hitting Set with Thin Sets} cannot be solved in time $2^{o(k \log{k})}$, unless ETH fails \cite{Daniel-2018}.
%\begin{theorem}\rm
%{\sc Offensive Alliance} can be solved  in $\mathcal{O}^{*}(2^{\mathcal{O}(k \log k)})$ time,  where $k$ is the solution size.
%\end{theorem}

%The algorithm of [?] is obtained by branching technique but it is not single-exponential. In this section  we show that in fact a single-exponential algorithm is very unlikely.

\begin{theorem}\label{o-klogk}
 Unless ETH fails, {\sc Offensive Alliance}  cannot be solved in time
 $\mathcal{O}^{*}(2^{o(k \log k)})$, where $k$ is  the solution size.
\end{theorem}

\proof We provide a polynomial-time algorithm that takes an
instance $(\mathcal{F},k)$ of $k \times k$ {\sc (Permutation) Hitting
Set with Thin Sets}, and outputs an  equivalent instance $(G,r)$ of
{\sc Offensive Alliance} with $r=5k$. We construct  $G$ the following way. 
\begin{enumerate}
    \item For 
every $F\in \mathcal{F}$, we introduce a vertex $v_F$ into $G$. Let 
$V_{\mathcal{F}}=\{v_F~|~F\in \mathcal{F}\}$.
  We also introduce 
a set   of $k^2$ vertices $W=\{w_{i,j}~:~ i\in [k], j\in [k]\}$. 
Make $v_F$ adjacent to $w_{ij}$ if $(i,j)\in F$.

\item We introduce a clique $D^{\triangle}$ of size $4k$ into $G$. 
For every $d \in D^{\triangle}$, we add a set of $10k$ vertices and make them adjacent to $d$.
For every $F\in \mathcal{F}$, we make $v_F$ adjacent to every vertex of $D^{\triangle}$.

\item We introduce another clique $D^{\square}$ of size $12k+1$ into $G$. 
Let $d_W(v_F)=d_F$. As we are dealing with thin sets, we have $d_F\leq k$ for
all $F\in \mathcal{F}$. For every 
$F\in \mathcal{F}$, we make $v_F$ adjacent to any $4k-d_F+1$ vertices of $D^{\square}$.

\item For every row $i\in [k]$, 
create a vertex $r_i$ into $G$ and make $r_i$ adjacent to $w_{ij}$ for all
$j\in [k]$. Let $R=\{r_1,\ldots,r_k\}$. For every $r\in R$, make $r$ adjacent to every vertex of $D^{\triangle}$ and 
any $3k+1$ vertices of $D^{\square}$.

\item For every column $j\in [k]$, 
create a vertex $c_j$ into $G$ and make $c_j$ adjacent to $w_{ij}$ for all
$i\in [k]$. Let $C=\{c_1,\ldots,c_k\}$. For every $c\in C$, make $c$ adjacent to every vertex of $D^{\triangle}$ and 
any $3k+1$ vertices of $D^{\square}$.
\end{enumerate}

\begin{figure}
    \centering
    \begin{tikzpicture}[scale=0.7]
\node[circle,draw,fill=black, inner sep=0 pt, minimum size=0.12cm]	(r51) at (.5,0.25) [label=above:]{}; 
\node[circle,draw,fill=black, inner sep=0 pt, minimum size=0.12cm]	(r52) at (1.8,0.25) [label=above:]{}; 
\node[circle,draw,fill=black, inner sep=0 pt, minimum size=0.12cm]	(r53) at (3.1,0.25) [label=above:]{}; 
\node[circle,draw,fill=black, inner sep=0 pt, minimum size=0.12cm]	(r54) at (4.4,0.25) [label=above:]{}; 
\node[circle,draw,fill=black, inner sep=0 pt, minimum size=0.12cm]	(r55) at (5.7,0.25) [label=above:]{};

\node[circle,draw,fill=black, inner sep=0 pt, minimum size=0.12cm]	(r41) at (.5,1.25) [label=above:]{}; 
\node[circle,draw,fill=black, inner sep=0 pt, minimum size=0.12cm]	(r42) at (1.8,1.25) [label=above:]{}; 
\node[circle,draw,fill=black, inner sep=0 pt, minimum size=0.12cm]	(r43) at (3.1,1.25) [label=above:]{}; 
\node[circle,draw,fill=black, inner sep=0 pt, minimum size=0.12cm]	(r44) at (4.4,1.25) [label=above:]{}; 
\node[circle,draw,fill=black, inner sep=0 pt, minimum size=0.12cm]	(r45) at (5.7,1.25) [label=above:]{};

\node[circle,draw,fill=black, inner sep=0 pt, minimum size=0.12cm]	(r31) at (.5,2.25) [label=above:]{}; 
\node[circle,draw,fill=black, inner sep=0 pt, minimum size=0.12cm]	(r32) at (1.8,2.25) [label=above:]{}; 
\node[circle,draw,fill=black, inner sep=0 pt, minimum size=0.12cm]	(r33) at (3.1,2.25) [label=above:]{}; 
\node[circle,draw,fill=black, inner sep=0 pt, minimum size=0.12cm]	(r34) at (4.4,2.25) [label=above:]{}; 
\node[circle,draw,fill=black, inner sep=0 pt, minimum size=0.12cm]	(r35) at (5.7,2.25) [label=above:]{};

\node[circle,draw,fill=black, inner sep=0 pt, minimum size=0.12cm]	(r21) at (.5,3.25) [label=above:]{}; 
\node[circle,draw,fill=black, inner sep=0 pt, minimum size=0.12cm]	(r22) at (1.8,3.25) [label=above:]{}; 
\node[circle,draw,fill=black, inner sep=0 pt, minimum size=0.12cm]	(r23) at (3.1,3.25) [label=above:]{}; 
\node[circle,draw,fill=black, inner sep=0 pt, minimum size=0.12cm]	(r24) at (4.4,3.25) [label=above:]{}; 
\node[circle,draw,fill=black, inner sep=0 pt, minimum size=0.12cm]	(r25) at (5.7,3.25) [label=above:]{};  

\node[circle,draw,fill=black, inner sep=0 pt, minimum size=0.12cm]	(r11) at (.5,4.25) [label=above:]{}; 
\node[circle,draw,fill=black, inner sep=0 pt, minimum size=0.12cm]	(r12) at (1.8,4.25) [label=above:]{}; 
\node[circle,draw,fill=black, inner sep=0 pt, minimum size=0.12cm]	(r13) at (3.1,4.25) [label=above:]{}; 
\node[circle,draw,fill=black, inner sep=0 pt, minimum size=0.12cm]	(r14) at (4.4,4.25) [label=above:]{}; 
\node[circle,draw,fill=black, inner sep=0 pt, minimum size=0.12cm]	(r15) at (5.7,4.25) [label=above:]{};

\draw[rounded corners,color=orange] (0, 0) rectangle (6.2, .5) {};  

\draw[rounded corners,color=orange] (0, 1) rectangle (6.2, 1.5) {};  

\draw[rounded corners,color=orange] (0, 2) rectangle (6.2, 2.5) {}; 

\draw[rounded corners,color=orange] (0, 3) rectangle (6.2, 3.5) {}; 

\draw[rounded corners,color=orange] (0, 4) rectangle (6.2, 4.5) {};

\draw[rounded corners,color=orange] (0, 5) rectangle (6.2, 5.5) {}; 

\draw[rounded corners,color=orange] (1.2, 7) rectangle (5, 7.5) {}; 

\draw[rounded corners,color=orange] (0.25, -0.25) rectangle (.75, 4.75) {}; 

\draw[rounded corners,color=orange] (1.55, -0.25) rectangle (2.05, 4.75) {}; 

\draw[rounded corners,color=orange] (2.85, -0.25) rectangle (3.35, 4.75) {}; 

\draw[rounded corners,color=orange] (4.15, -0.25) rectangle (4.65, 4.75) {}; 

\draw[rounded corners,color=orange] (5.45, -0.25) rectangle (5.95, 4.75) {}; 

\draw[rounded corners,color=orange] (-0.25, -0.25) rectangle (-.7, 4.75) {}; 

\draw[rounded corners,color=orange] (0, 5.5) rectangle (6.2, 5) [label=left:]{};

\node[square]	(r5) at (-.5,0.25) [label=left:$r_{5}$]{}; 
\node[square]	(r4) at (-.5,1.25) [label=left:$r_{4}$]{}; 
\node[square]	(r3) at (-.5,2.25) [label=left:$r_{3}$]{}; 
\node[square]	(r2) at (-.5,3.25) [label=left:$r_{2}$]{}; 
\node[square]	(r1) at (-.5,4.25) [label=left:$r_{1}$]{}; 

\node[]	(r05) at (0.1,0.25) [label=left:]{}; 
\node[]	(r04) at (0.1,1.25) [label=left:]{}; 
\node[]	(r03) at (0.1,2.25) [label=left:]{}; 
\node[]	(r02) at (0.1,3.25) [label=left:]{}; 
\node[]	(r01) at (0.1,4.25) [label=left:]{};

\node[square]	(c1) at (.5,5.25) [label=right:$c_{1}$]{}; 
\node[square]	(c2) at (1.8,5.25) [label=right:$c_{2}$]{}; 
\node[square]	(c3) at (3.1,5.25) [label=right:$c_{3}$]{}; 
\node[square]	(c4) at (4.4,5.25) [label=right:$c_{4}$]{}; 
\node[square]	(c5) at (5.7,5.25) [label=right:$c_{5}$]{};   

\node[]	(c01) at (.5,4.65) [label=right:]{}; 
\node[]	(c02) at (1.8,4.65) [label=right:]{}; 
\node[]	(c03) at (3.1,4.65) [label=right:]{}; 
\node[]	(c04) at (4.4,4.65) [label=right:]{}; 
\node[]	(c05) at (5.7,4.65) [label=right:]{}; 

\draw[thick](c1)--(c01);
\draw[thick](c2)--(c02);
\draw[thick](c3)--(c03);
\draw[thick](c4)--(c04);
\draw[thick](c5)--(c05);

\draw[thick](r1)--(r01);
\draw[thick](r2)--(r02);
\draw[thick](r3)--(r03);
\draw[thick](r4)--(r04);
\draw[thick](r5)--(r05);

\node[](C00) at (6.2,5.25) [label=right:$C$]{};
\node[](R00) at (-1.2,2.5) [label=left:$R$]{};
\node[](W00) at (6.2,0.5) [label=right:$W$]{};

\node[square]	(f1) at (1.8,7.25) [label=above:$v_{F_{1}}$]{}; 
\node[square]	(f2) at (3.1,7.25) [label=above:$v_{F_{2}}$]{}; 
\node[square]	(f3) at (4.4,7.25) [label=above:$v_{F_{3}}$]{};

\draw[green](f1)--(r11);
\draw[green](f1)--(r21);
\draw[green](f1)--(r44);
\draw[green](f1)--(r53);

\draw[green](f2)--(r14);
\draw[green](f2)--(r34);
\draw[green](f2)--(r51);

\draw[green](f3)--(r11);
\draw[green](f3)--(r25);
\draw[green](f3)--(r32);
\draw[green](f3)--(r55);

\node[circle,draw, inner sep=0 pt, minimum size=0.8cm]	(ds) at (-.4,6.5) [label=above:]{$D^{\square}$};  

\node[circle,draw, inner sep=0 pt, minimum size=0.8cm]	(dt) at (-1.5,7.5) [label=above:]{$D^{\triangle}$}; 

\node[circle,draw,fill=black, inner sep=0 pt, minimum size=0.1cm]	(x) at (-1.5,7.8) [label=above:]{};

\node[circle,draw,fill=black, inner sep=0 pt, minimum size=0.1cm]	(x1) at (-1.7,8.6) [label=above:]{};
\node[circle,draw,fill=black, inner sep=0 pt, minimum size=0.1cm]	(x2) at (-1.5,8.6) [label=above:]{};
\node[circle,draw,fill=black, inner sep=0 pt, minimum size=0.1cm]	(x3) at (-1.1,8.6) [label=above:]{};

\draw(x)--(x1);
\draw(x)--(x2);
\draw(x)--(x3);
\draw[thick,dotted](x2)--(x3);

\node[]	(f) at (1.3,7.25) [label=right:]{};
\node[]	(r) at (-.5,4.6) [label=right:]{};
\node[]	(c) at (0.1,5.25) [label=right:]{};

\draw[thick](dt)--(f);
\draw[thick](ds)--(f);
\draw[thick](dt).. controls (-1.5,5.25).. (c);
\draw[thick](ds)--(c);
\draw[thick](dt)--(r);
\draw[thick](ds)--(r);

    \end{tikzpicture}
    \caption{Example of the reduction in Theorem \ref{o-klogk}
    applied to an instance $(\mathcal{F}=\{F_1,F_2,F_3\},5)$ of 
    $5\times 5$ {\sc Permutation Hitting Set with Thin Sets} 
    that has three sets  $F_{1}=\{(1,1),(2,1),(4,4),(5,3)\}, F_{2}=\{(1,4),(3,4),(5,1)\},F_{3}=\{(1,1),(2,5),(3,2),(5,5)\}$ 
    with each set containing at most one element from each row.}
    \label{solsizelowerbound}
\end{figure}
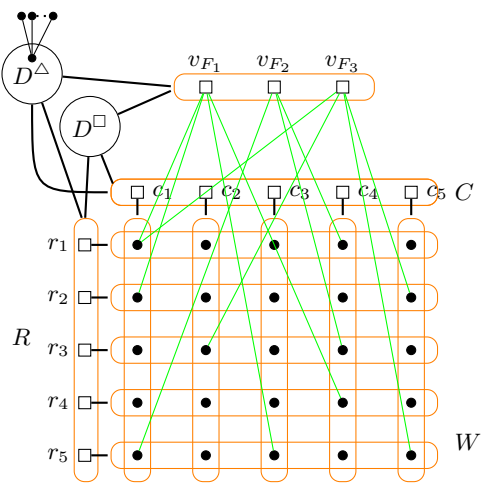

This completes the construction of $G$. Set $r=5k$.
 We now formally argue that
instances $(\mathcal{F},k)$ and $(G,r)$ are  equivalent. 
Assume 
first that $X$ is a solution to the instance $(\mathcal{F},k)$. 
For each $i\in [k]$, let $j_i$ be the unique index such that 
$(i,j_i) \in X$. We claim that the set $$S= D^{\triangle} \cup
\{w_{1j_1}, w_{2j_2},\ldots,w_{kj_k}\}$$ is an offensive alliance of size exactly $5k$ in $G$. 
We see that $N(S)= R \cup C \cup  \{v_F ~:~ F\in \mathcal{F}\}$.
Let $v$ be an arbitrary element of $N(S)$. We need to prove that $d_S(v)\geq d_{S^c}(v)+1$ for all $v\in N(S)$.
If $v$ is an element of  $R$ or $C$, the neighbours of $v$ in $S$ are the 
elements of $D^{\triangle}$ and one element from $W$. Thus we have $d_S(v) =4k+1$.
The neighbours of $v$ in $S^c$ are $3k+1$ elements of $D^{\square}$
and $k-1$ elements of $W$; therefore we have $d_{S^{c}}(v) = 4k$. 
If $v$ is an element of $\{v_F ~:~ F\in \mathcal{F}\}$, the neighbours of $v$ in $S$ are 
the 
elements of $D^{\triangle}$ and at least one element from $W$ as $X$ is a hitting set; 
thus we have $d_S(v)\geq 4k+1$. The neighbours of $v$ in $S^c$ are 
$4k-d+1$ 
elements of $D^{\square}$ and at most $d-1$ elements from $W$; 
thus we have $d_{S^c}(v)\leq (4k-d+1)+(d-1)= 4k$.
 This shows that $S$ is indeed an offensive alliance.

\par In the reverse direction, let $S$ be an offensive alliance of size at most $5k$ in $G$. 
First we show that it can be assumed that $  N(S) \cap D^{\square}= \emptyset$.  Suppose, for the sake of contradiction, that 
$v\in N(S)\cap D^{\square}$. Then $v$ must satisfy the condition $d_S(v)\geq d_{S^c}(v)+1$. 
As $v$ has degree at least $12k$, in order to satisfy the condition $d_S(v)\geq d_{S^c}(v)+1$,
the size of $S$ must be at least $6k$, a contradiction to the assumption that 
the size of $S$ is at most $5k$.  Next we show that it can be assumed 
that $ S \cap D^{\square}= \emptyset$.
Suppose that $S$ contains an element of $D^{\square}$.
As $D^{\square}$ is a clique and  $D^{\square} \cap N(S)=\emptyset$, if $S$ contains one element of $D^{\square}$ then 
$D^{\square}\subseteq S$.  This is not possible as $D^{\square}$ has $12k+1$
elements and $S$ has at most $5k$ elements. 
Therefore, we may assume that $ (S\cup N(S)) \cap D^{\square}= \emptyset$.
This in turn implies that that $(R \cup C \cup V_{\mathcal{F}}) \cap S = \emptyset$. 
 %As all the vertices in $D^{\triangle}$ have degree at least $14k$, it implies that $D^{\triangle} \cap N(S)= \emptyset$. 
As the offensive alliance $S$ is non-empty, we have 
$S \cap D^{\triangle}\neq \emptyset $ or $S \cap W \neq \emptyset$. 
Note that in either case, we get 
$N(S) \cap (R\cup C \cup V_{\mathcal{F}}) \neq \emptyset$. 
Let $u$ be an arbitrary element of $ N(S) \cap (R \cup C \cup V_{\mathcal{F}})$. 
If $S \cap D^{\triangle} = \emptyset$ then clearly we have 
$d_{S}(u) < d_{S^{c}}(u)+1$ which is not possible. 
Therefore $S \cap D^{\triangle} \neq \emptyset$. We observe  that if $S$ contains one element of $D^{\triangle}$ then 
it contains all elements of $D^{\triangle}$, that is,  $D^{\triangle} \subseteq S$. 
As $D^{\triangle} \subseteq S$ and $(R \cup C \cup V_{\mathcal{F}}) \cap S=\emptyset$, we get $(R \cup C \cup V_{\mathcal{F}}) \subseteq N(S)$. 
As $S$ is an offensive alliance, every element $u$ of $N(S)$ has to satisfy
the condition $d_{S}(u) \geq d_{S^{c}}(u)+1$.
Consider an arbitrary vertex $r_i$ of $R$. 
If $S \cap \{w_{ij}~:~j\in[k]\} = \emptyset$ then $d_{S}(r_{i})=4k$ and  $d_{S^{c}}(r_{i})=4k$ which is not possible as $r_i$ does not satisfy  the condition $d_{S}(r_{i}) 
\geq  d_{S^{c}}(r_{i})+1$. This implies that, for each $i\in [k]$,
$S$ contains at least one element from $\{w_{ij}~:~j\in [k]\}$ but since
$|S|\leq 5k$ and $D^{\triangle} \subseteq S$, 
 $S$ contains exactly one element from $\{w_{ij}~:~j\in [k]\}$. 
 Using the same argument for an arbitrary vertex $c_j\in C$, we get that 
 $S$ contains exactly one element from $\{w_{ij}~:~i\in [k]\}$.  We claim that 
$X= \{(i,j)~|~ w_{ij} \in S\}$ is a permutation hitting set of size $k$. 
Let us assume that there exists a set $F\in \mathcal{F}$ which is not hit by 
$X$. In that case, we have $d_{S}(v_F)=4k$ and $d_{S^{c}}(v_F) = (4k - d_F)+ d_F = 4k$. This means that $d_{S}(v_F) < d_{S^{c}}(v_F)+1$ which is a contradiction. Therefore, 
 $X$ is a hitting set of size $k$. As $X$ has exactly one element in 
 each row and in each column, $X$ is a permutation hitting set.  

An algorithm solving {\sc Offensive Alliance} in time $2^{o(k \log k)}$ would therefore translate into an algorithm running in time $2^{o(k \log k)}$ for $k \times k$ {\sc (Permutation) Hitting
Set with Thin Sets} and contradicts the ETH. \qed\\

\begin{corollary}
 Unless ETH fails, {\sc Exact Offensive Alliance} problem cannot be solved in time $\mathcal{O}^{*}(2^{o(k \log k)})$, where $k$ is the solution size.
\end{corollary}

\section{No polynomial kernel parameterized by solution size and vertex cover}
Parameterized by the solution size, the problem is FPT  and in this section we show the following kernelization hardness of {\sc Offensive Alliance}. 
 
 \begin{theorem}\label{No-Poly}\rm
{\sc Offensive Alliance}  parameterized by the solution size and vertex cover combined does not admit 
a polynomial compression unless coNP $\subseteq$ NP/poly.
 \end{theorem}
  \noindent To prove Theorem \ref{No-Poly}, we give a polynomial parameter transformation 
  (PPT) from {\sc Closest String} to {\sc Offensive Alliance} parameterized by
  the solution size. In the  {\sc Closest string} problem, we are given a set of $k$ strings 
  $\mathcal{X} = \{x_{1}, x_2,\ldots,x_{k}\}$, each string  over an alphabet $\Sigma$ and of length
  $n$, and an integer $d$. 
The objective is to check whether there exists a string $y$ of length $n$ over $\Sigma$
such that $d_{H}(y, x_{i}) \leq d$ for all $i \in \{1,2,\ldots,k\}$. Here
$d_{H}(x,y)$ is the \emph{Hamming distance} between strings $x$ and $y$, that is, 
 the
number of places where strings $x$ and $y$ differ. We call any such string $y$
a \emph{central string}.
Let $x$ be a string over alphabet $\Sigma$. We denote the letter on
the $p$-th position of $x$ as $x[p]$. Thus $x=x[1]x[2]\ldots x[n]$ for a string 
of length $n$. We say string $x$ and $y$ \emph{differ} on the $p$-th position if 
$x[p]\neq y[p]$.
The following theorem is known:
\begin{theorem}\cite{Basavaraju2018OnTK}
$(d,n)${\sc Closest String} does not admit a polynomial kernel unless NP $\subseteq$ coNP/poly.
\end{theorem}

\noindent They also observe that the kernelization lower bound for  {\sc Closest String} works for any fixed alphabet $\Sigma$ of size at least two. 
Therefore, without loss of generality, we assume that $\Sigma =\{A_{1},A_{2}\}$.

 \subsection{Proof of Theorem \ref{No-Poly}} 
%Here $s_{i}[j]$ denotes the $j$th entry of string $s_{i}$. 

We give a PPT from the  {\sc Closest String} problem. 
Given an instance $(\mathcal{X},d)$ of the {\sc Closest String} problem, 
we construct an instance   $(G,r)$ of {\sc Offensive Alliance} the following way. 
\begin{enumerate}
    \item For 
every $x\in \mathcal{X}$, we introduce a vertex $v_x$ into $G$. Let 
$V_{\mathcal{X}}=\{v_x~|~x\in \mathcal{X}\}$.
  We also introduce 
a set   of $2n$ vertices $W=\{w_{i,j}~:~ i\in [n], j\in [2]\}$. 
Make $v_x$ adjacent to $w_{i1}$ if the letter on the $i$th position of $x$ is 
$A_1$; make $v_x$ adjacent to $w_{i2}$ if the letter on the $i$th position of $x$ is 
$A_2$.

\item We introduce a clique $D^{\triangle}$ of size $3n+2d+1$ into $G$. 
For every $d \in D^{\triangle}$, we add a set $V_d$ of $12n$ vertices and 
make them adjacent to $d$.
For every $v_x\in V_{\mathcal{X}}$, we make $v_x$ adjacent to every vertex of $D^{\triangle}$.

\item We introduce another clique $D^{\square}$ of size $12n+1$ into $G$. 
 For every $v_x \in V_{\mathcal{F}}$, we make $v_x$ adjacent to any $4n$ vertices of $D^{\square}$.

\item For every row $i\in [n]$, 
create a vertex $r_i$ into $G$ and make $r_i$ adjacent to $w_{i1}$ and $w_{i2}$. 
Let $R=\{r_1,\ldots,r_n\}$. For every $r\in R$, make $r$ adjacent to 
any three vertices of $D^{\triangle}$ and 
any two vertices of $D^{\square}$.
\end{enumerate}

\begin{figure}
    \centering
   \begin{tikzpicture}[scale=0.7]

\draw[rounded corners,color=orange] (0, 0) rectangle (3,0.5) {};
   
\node[circle,draw,fill=black, inner sep=0 pt, minimum size=0.12cm]	(x1) at (0.5,0.25) [label=above:$v_{x_{1}}$]{};   
\node[circle,draw,fill=black, inner sep=0 pt, minimum size=0.12cm]	(x2) at (1.5,0.25) [label=above:$v_{x_{2}}$]{};      
\node[circle,draw,fill=black, inner sep=0 pt, minimum size=0.12cm]	(x3) at (2.5,0.25) [label=above:$v_{x_{3}}$]{};     
   
\node[circle,draw,fill=black, inner sep=0 pt, minimum size=0.12cm]	(a1) at (-1,-2) [label=above:]{};        
 \node[circle,draw,fill=black, inner sep=0 pt, minimum size=0.12cm]	(b1) at (4,-2) [label=above:]{};   
  \node[circle,draw,red, inner sep=0 pt, minimum size=0.22cm]	(u1) at (4,-2) [label=above:]{};  
   
 \node[circle,draw,fill=black, inner sep=0 pt, minimum size=0.12cm]	(a2) at (-1,-2.8) [label=above:]{};        
 \node[circle,draw,fill=black, inner sep=0 pt, minimum size=0.12cm]	(b2) at (4,-2.8) [label=above:]{};   
 \node[circle,draw,red, inner sep=0 pt, minimum size=0.22cm]	(u2) at (-1,-2.8) [label=above:]{};

 \node[circle,draw,fill=black, inner sep=0 pt, minimum size=0.12cm]	(a3) at (-1,-3.6) [label=above:]{};        
 \node[circle,draw,fill=black, inner sep=0 pt, minimum size=0.12cm]	(b3) at (4,-3.6) [label=above:]{};    
  \node[circle,draw,red, inner sep=0 pt, minimum size=0.22cm]	(u3) at (-1,-3.6) [label=above:]{}; 
   
 \node[circle,draw,fill=black, inner sep=0 pt, minimum size=0.12cm]	(a4) at (-1,-4.4) [label=above:]{};        
 \node[circle,draw,fill=black, inner sep=0 pt, minimum size=0.12cm]	(b4) at (4,-4.4) [label=above:]{};    
   \node[circle,draw,red, inner sep=0 pt, minimum size=0.22cm]	(u4) at (-1,-4.4) [label=above:]{};

 \node[circle,draw,fill=black, inner sep=0 pt, minimum size=0.12cm]	(a5) at (-1,-5.2) [label=above:]{};        
 \node[circle,draw,fill=black, inner sep=0 pt, minimum size=0.12cm]	(b5) at (4,-5.2) [label=above:]{};  
\node[circle,draw,red, inner sep=0 pt, minimum size=0.22cm]	(u5) at (-1,-5.2) [label=above:]{}; 
 
 \node[circle,draw,fill=black, inner sep=0 pt, minimum size=0.12cm]	(a6) at (-1,-6.0) [label=above:]{};        
 \node[circle,draw,fill=black, inner sep=0 pt, minimum size=0.12cm]	(b6) at (4,-6.0) [label=above:]{};   
 \node[circle,draw,red, inner sep=0 pt, minimum size=0.22cm]	(u6) at (-1,-6.0) [label=above:]{}; 
   
 \node[circle,draw,fill=black, inner sep=0 pt, minimum size=0.12cm]	(a7) at (-1,-6.8) [label=above:]{};        
 \node[circle,draw,fill=black, inner sep=0 pt, minimum size=0.12cm]	(b7) at (4,-6.8) [label=above:]{};  
 \node[circle,draw,red, inner sep=0 pt, minimum size=0.22cm]	(u7) at (-1,-6.8) [label=above:]{};
 
\node[circle,draw,fill=black, inner sep=0 pt, minimum size=0.12cm]	(r1) at (-2.5,-2) [label=above:]{};   
\node[circle,draw,fill=black, inner sep=0 pt, minimum size=0.12cm]	(r2) at (-2.5,-2.8) [label=above:]{};   
\node[circle,draw,fill=black, inner sep=0 pt, minimum size=0.12cm]	(r3) at (-2.5,-3.6) [label=above:]{};   
\node[circle,draw,fill=black, inner sep=0 pt, minimum size=0.12cm]	(r4) at (-2.5,-4.4) [label=above:]{};   
\node[circle,draw,fill=black, inner sep=0 pt, minimum size=0.12cm]	(r5) at (-2.5,-5.2) [label=above:]{};   
\node[circle,draw,fill=black, inner sep=0 pt, minimum size=0.12cm]	(r6) at (-2.5,-6) [label=above:]{};   
\node[circle,draw,fill=black, inner sep=0 pt, minimum size=0.12cm]	(r7) at (-2.5,-6.8) [label=above:]{};

\draw(r1)--(a1);
\draw(r2)--(a2);
\draw(r3)--(a3);
\draw(r4)--(a4);
\draw(r5)--(a5);
\draw(r6)--(a6);
\draw(r7)--(a7);

\draw(r1)..controls(-1,-1.7)..(b1);
\draw(r2)..controls(-1,-2.5)..(b2);
\draw(r3)..controls(-1,-3.3)..(b3);
\draw(r4)..controls(-1,-4.1)..(b4);
\draw(r5)..controls(-1,-4.9)..(b5);
\draw(r6)..controls(-1,-5.7)..(b6);
\draw(r7)..controls(-1,-6.5)..(b7);

\draw[rounded corners,color=orange] (-2.75, -1.5) rectangle (-2.25,-7.3) {};

\draw[green](x1)--(b1);
\draw[green](x1)--(a2);
\draw[green](x1)--(b3);
\draw[green](x1)--(b4);
\draw[green](x1)--(b5);
\draw[green](x1)--(a6);
\draw[green](x1)--(a7);

\draw[green](x2)--(b1);
\draw[green](x2)--(b2);
\draw[green](x2)--(a3);
\draw[green](x2)--(b4);
\draw[green](x2)--(a5);
\draw[green](x2)--(b6);
\draw[green](x2)--(a7);

\draw[green](x3)--(b1);
\draw[green](x3)--(b2);
\draw[green](x3)--(a3);
\draw[green](x3)--(a4);
\draw[green](x3)--(a5);
\draw[green](x3)--(a6);
\draw[green](x3)--(b7);

\node[circle,draw, inner sep=0 pt, minimum size=1.2cm]	(ds) at (-3.5,0) [label=above:]{$D^{\square}$};  

\node[circle,draw, inner sep=0 pt, minimum size=1.2cm]	(dt) at (-2.5,1.5) [label=above:]{$D^{\triangle}$}; 

\node[circle,draw,fill=black, inner sep=0 pt, minimum size=0.1cm]	(x) at (-2.5,1.9) [label=left:$d$]{};
\node[circle,draw,fill=black, inner sep=0 pt, minimum size=0.1cm]	(x1) at (-2.7,2.7) [label=above:]{};
\node[circle,draw,fill=black, inner sep=0 pt, minimum size=0.1cm]	(x2) at (-2.5,2.7) [label=above:$V_{d}$]{};
\node[circle,draw,fill=black, inner sep=0 pt, minimum size=0.1cm]	(x3) at (-2.1,2.7) [label=above:]{};

\draw(x)--(x1);
\draw(x)--(x2);
\draw(x)--(x3);
\draw[thick,dotted](x2)--(x3);

\node[]	(f) at (0.15,.25) [label=right:]{};
\node[]	(r) at (-2.5,-1.65) [label=right:]{};

\node[]	(vx) at (3.15,.25) [label=right:$V_{\mathcal{X}}$]{};
\node[]	(R11) at (-3,-4.4) [label=left:$R$]{};
\node[]	(W11) at (5,-4.4) [label=left:$W$]{};

\node[]	(t1) at (-2.5,-1.05) [label=right:$3$]{};
\node[]	(t2) at (-3.5,-1.05) [label=right:$2$]{};

\draw[thick](ds)--(f);
\draw[thick](dt)--(f);
\draw[thick](ds)--(r);
\draw[thick](dt)--(r);

   \end{tikzpicture}
    \caption{Example of reduction of Theorem \ref{No-Poly} applied to an instance $(\chi, d)$ of {\sc Closest String} where $\chi$ contains three strings $x_{1}=1011100, x_{2}=1101010,x_{3}=1110001$ and $d=3$. A solution string $y=1000000$ is shown in red circles.}
    \label{Nopolykernel}
\end{figure}
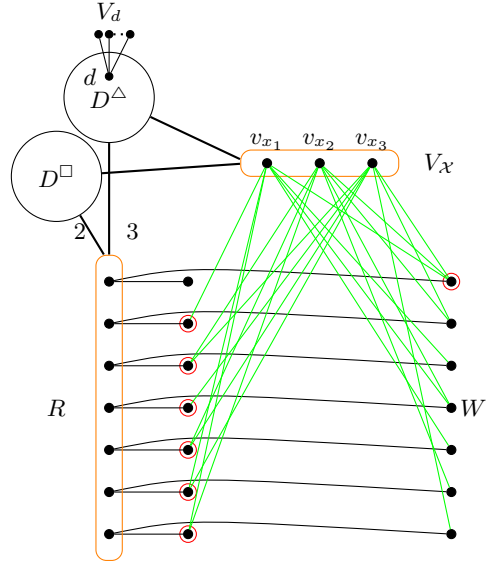

This completes the construction of $G$. Note that the set $R \cup W \cup D^{\square} \cup D^{\triangle}$ forms a vertex cover of $G$ of size $18n+2d+2$. We set $r= 4n+2d+1$. It is easy to see that the above construction takes polynomial time. 
%Next, we prove that there exists a string $y$ such that $d_{H}(x_i,y) \leq d$ for all $i \in \{1,2,\ldots,k\}$ if and only if there exists an offensive alliance of size at most $r$ in $G$.
We now formally argue that instances $(\mathcal{X},d)$ and $(G,r)$ are  equivalent. 
Assume first that $y$ is a solution to the instance $(\mathcal{X},d)$, that is, 
$d_{H}(x_i,y) \leq d$ for all $i \in \{1,2,\ldots,k\}$. 
We claim that 
$$S= D^{\triangle} \cup \Big\{w_{ij} ~|~ y[i]=A_j ~\text{for}~ i=1,2,\ldots,n\Big\} %\bigcup\limits_{d\in D^{\triangle} }{V_d}
$$ 
is an offensive alliance of size at most $r$. 
We see that $N(S) =R\cup V_{\mathcal{X}}$. 
We need to prove that $d_S(v)\geq d_{S^c}(v)+1$ for every 
$v\in R\cup V_{\mathcal{X}}$. If $r$ is an element of  $R$, 
the neighbours of $r$ in $S$ are three 
elements of $D^{\triangle}$ and one element from $W$. Thus we have $d_S(r) =4$.
The neighbours of $r$ in $S^c$ are two elements of $D^{\square}$
and one elements of $W$; therefore we have $d_{S^{c}}(r) = 3$ and $r$ satisfies the 
required condition. If $v_x$ is an element of $V_{\mathcal{X}}$, the neighbours of $v_x$ in 
$S$ are $3n+2d+1$ elements of $D^{\triangle}$ and $n-d_H(x,y)$ element from $W$. 
Thus we have $d_S(v_x) \geq  (3n+2d+1)+(n-d)=4n+d+1$. 
The neighbours of $v_x$ in 
$S^c$ are $4n$ elements of $D^{\square}$ and $d_H(x,y)$ element from $W$; hence
 $d_{S^c}(v_x)= 4n+d_H(x,y)\leq 4n+d$. Therefore $v_x$ satisfies the required 
 condition. 

\par In the reverse direction, let $S$ be an offensive alliance of size at most $4n+2d+1$ in $G$. First we show that it can be assumed that $  N(S) \cap D^{\square}= \emptyset$.  Suppose, for the sake of contradiction, that 
$v\in N(S)\cap D^{\square}$. Then $v$ must satisfy the condition $d_S(v)\geq d_{S^c}(v)+1$. 
As $v$ has degree at least $12k$, in order to satisfy the condition $d_S(v)\geq d_{S^c}(v)+1$,
the size of $S$ must be at least $6k$, a contradiction to the assumption that 
the size of $S$ is at most $4n+2d+1$. Next we show that it can be assumed 
that $ S \cap D^{\square}= \emptyset$.
Suppose that $S$ contains an element of $D^{\square}$.
As $D^{\square}$ is a clique and  $D^{\square} \cap N(S)=\emptyset$, 
if $S$ contains one element of $D^{\square}$ then 
$D^{\square}\subseteq S$.  This is not possible as $D^{\square}$ has $12n+1$
elements and $S$ has at most $4n+2d+1$ elements. 
Therefore, we may assume that $ (S\cup N(S)) \cap D^{\square}= \emptyset$.
This in turn implies  that $S$ does not contain any element from $V_{\mathcal{X}}\cup R$. 
As the offensive alliance $S$ is non-empty, we have 
$S \cap D^{\triangle}\neq \emptyset $ or $S \cap W \neq \emptyset$. 
Note that in either case, we get 
$N(S) \cap (R\cup V_{\mathcal{X}}) \neq \emptyset$. Let $u$ be an arbitrary element of $ N(S) \cap (R\cup V_{\mathcal{X}})$. 
If $S \cap D^{\triangle} = \emptyset$ then clearly we have 
$d_{S}(u) < d_{S^{c}}(u)+1$ which is not possible. 
Therefore $S \cap D^{\triangle} \neq \emptyset$. We observe  that if $S$ contains one element of $D^{\triangle}$ then 
it contains all elements of $D^{\triangle}$, that is,  $D^{\triangle} \subseteq S$. 
As $D^{\triangle} \subseteq S$ and $(R  \cup V_{\mathcal{X}}) \cap S=\emptyset$, we get $(R\cup V_{\mathcal{X}}) \subseteq N(S)$. 
As $S$ is an offensive alliance, every element $u$ of $N(S)$ has to satisfy
the condition $d_{S}(u) \geq d_{S^{c}}(u)+1$.
Consider an arbitrary vertex $r_i$ of $R$. 
If $S \cap \{w_{i1}, w_{i2} \} = \emptyset$ then $d_{S}(r_{i})=3$ and  
$d_{S^{c}}(r_{i})=4$ which is not possible as $r_i$ does not satisfy  the condition $d_{S}(r_i) 
\geq  d_{S^{c}}(r_{i})+1$. This implies that, for each $i\in [n]$,
$S$ contains at least one element from $\{w_{i1}, w_{i2}\}$.  Since
$|S|\leq 4n+2d+1$ and $D^{\triangle} \subseteq S$, 
 $S$ contains exactly one element from $\{w_{i1},w_{i2}\}$ for each $i$. 
Define a string $y=y[1]y[2]\ldots y[n]$, where $y[i]=1$ if $w_{i1}\in S$ and 
$y[i]=2$ if $w_{i2}\in S$. We claim $y$ is a central string. Assume, for the sake of contradiction, that there exists a string $x_i$ such that $d_{H}(x_i,y)>d$. 
In this case, we see that $d_S(v_{x_i})< (3n+2d+1) + n-d \leq 4n+d$ and 
$d_{S^c}(v_{x_i}) > 4n + d$. Therefore, $d_S(v_{x_i}) < d_{S^c}(v_{x_i})+1$, which is a contradiction. \qed\\

\section{Faster FPT algorithms parameterized by vertex cover number}
We know that both  {\sc Offensive Alliance} and {\sc Strong Offensive Alliance} admit  FPT algorithms \cite{KIYOMI201791} when parameterized by the vertex cover number of the input graph. The algorithms in \cite{KIYOMI201791} use {\sc Integer Linear Programming}, and thus their dependency on the parameter may be gigantic. The reason is this.
 The {\sc Offensive Alliance} problem is mapped to  an ILP with at most $2^{\tt{vc(G)}}$ many variables where $\tt{vc(G)}$ is the vertex cover number of the input graph. It is proved in \cite{fellows} that {\sc $p$-Variable Integer Linear Programming Optimization ($p$-Opt-ILP)} can be solved using $O(p^{2.5p+o(p)}\cdot L \cdot log(MN))$ arithmetic operations and space polynomial in $L$. Thus the algorithm in \cite{KIYOMI201791} requires $\mathcal{O}^{*}({(2^{\tt{vc(G)}})}^{\mathcal{O}(2^{\tt{vc(G)})}})$ time. 
The natural question would be whether they admit $\mathcal{O}^{*}(\tt{vc(G)}^{\mathcal{O}(\tt{vc(G)}})$ time algorithm. We answer this question with the following theorem.

\begin{theorem}\rm \label{vc-improved}
 {\sc Offensive Alliance}  can be solved in time $\mathcal{O}^{*}(\tt{vc(G)}^{\mathcal{O}(\tt{vc(G)}})$ where $\tt{vc(G)}$ is the vertex cover number of the input graph $G$.
\end{theorem}
\proof Let $C$ be a vertex cover of $G$ of size {\tt{vc(G)}}.
Note that  $C$ forms an offensive alliance. This is because $N(C)$ is an independent set and 
every vertex of $N(C)$ has no neighbours in $C^c$ and has at least one neighbour in $C$. 
Therefore we have $d_C(v)\geq d_{C^c}(v)+1$ for all $v\in N(C)$, and hence $C$ is an offensive alliance.
This implies that the size of minimum offensive alliance is at most $\tt{vc(G)}$. In \cite{DA-FPT}, it was proved using branching technique that {\sc Offensive Alliance} problem parameterized by solution size admits a $\mathcal{O}(n^{2}k(2k)^{k-1})$. This implies that, we have an algorithm with running time $\mathcal{O}^{*}(\tt{vc(G)})^{\tt{vc(G)}}$. \qed\\\\

\noindent  The arguments in the proof of Theorem \ref{vc-improved} is  also applicable to {\sc Strong Offensive Alliance} as long as the input graph $G$ has minimum degree at least two.
As a direct consequence of Theorem \ref{vc-improved}, we have the following corollary.
\begin{corollary}\rm
  {\sc Strong Offensive Alliance}  can be solved in time $\mathcal{O}^{*}(\tt{vc(G)}^{\mathcal{O}(\tt{vc(G)}})$ where $\tt{vc(G)}$ is the vertex cover number of the input graph $G$ with $\delta(G)\geq 2$.
\end{corollary}

\section{Classical lower bounds under ETH}

%\subsection{ {\sc Offensive Alliance} on general graphs}

%In this section, we give  an algorithmic lower bound for the
%{\sc Defensive Alliance} problem using ETH. 
%The Exponential Time Hypothesis (ETH) is a conjecture stating that,
%roughly speaking, $n$-variable 3-SAT cannot be solved in time $2^{o(n)}$. 
In this section, we prove lower bound based on ETH
for the time needed to solve the {\sc Offensive Alliance} problem. 
In order to prove that a too fast algorithm for {\sc Offensive Alliance} contradicts ETH, we give a reduction from {\sc Vertex Cover} in graphs of maximum degree 3
 and argue that a too fast algorithm for 
{\sc Offensive Alliance} would solve {\sc Vertex Cover} in graphs of maximum degree 3
 in time $2^{o(n)}$.  Johnson  and Szegedy \cite{soda1999} proved that, assuming ETH, there is no algorithm with running time $2^{o(n)}$ to compute a minimum vertex cover in graphs of maximum degree 3.

\begin{figure}[ht]
    \centering
   \begin{tikzpicture}[scale=0.7]
\draw [draw=green](-4,-1.5) ellipse (.7cm and 2.5cm); 
\draw [draw=green](-6,-1.5) ellipse (.7cm and 2.5cm); 
\draw [draw=green](-8,-1.5) ellipse (.7cm and 2.5cm); 
\node (E0) at (-4,1) [label=above:${E_0}$] {};
\node (E0) at (-6,1) [label=above:${V_0}$] {};
\node (E0) at (-8,1) [label=above:${V_1}$] {};
\node[circle, draw, inner sep=0pt, minimum size=5pt] (v1) at (-16,-1.5) [label=above:${v_{1}}$] {};
\node[circle, draw, inner sep=0pt, minimum size=5pt] (v2) at (-17,-3) [label=below:${v_{2}}$] {};
\node[circle, draw, inner sep=0pt, minimum size=5pt] (v3) at (-15,-3) [label=below:${v_{3}}$] {};

\node[circle, draw, inner sep=0pt, minimum size=5pt] (v01) at (-6,0) [label=below:${v^0_{1}}$] {};
\node[circle, draw, inner sep=0pt, minimum size=5pt] (v02) at (-6,-1.5) [label=below:${v^0_{2}}$] {};
\node[circle, draw, inner sep=0pt, minimum size=5pt] (v03) at (-6,-3) [label=below:${v^0_{3}}$] {};

\node[circle, draw, inner sep=0pt, minimum size=5pt] (v11) at (-8,0) [label=below:${v^1_{1}}$] {};
\node[circle, draw, inner sep=0pt, minimum size=5pt] (v12) at (-8,-1.5) [label=below:${v^1_{2}}$] {};
\node[circle, draw, inner sep=0pt, minimum size=5pt] (v13) at (-8,-3) [label=below:${v^1_{3}}$] {};

\node[circle, draw, inner sep=0pt, minimum size=5pt] (e1) at (-4,0) [label=below:$e_1$] {};
\node[circle, draw, inner sep=0pt, minimum size=5pt] (e2) at (-4,-1.5) [label=below:$e_2$] {};
\node[circle, draw, inner sep=0pt, minimum size=5pt] (e3) at (-4,-3) [label=below:$e_3$] {};

\node[circle, draw, inner sep=0pt, minimum size=5pt] (a) at (-2,-0.5) [label=below:$a$] {};

\node[circle, draw, inner sep=0pt, minimum size=5pt] (t) at (-2,-2.5) [label=below:$e$] {};
%\node[circle, draw, inner sep=0pt, minimum size=5pt] (t1) at (-1,-2.5) [] {};
\node[circle, draw, inner sep=0pt, minimum size=5pt] (t2) at (-1,-3) [] {};
\node[circle, draw, inner sep=0pt, minimum size=5pt] (t3) at (-1,-2) [] {};
\draw[dotted, thick] (-1,-2.3) -- (-1,-2.7);

\node[circle, draw, inner sep=0pt, minimum size=5pt] (a1) at (-1,-1) [] {};
%\node[circle, draw, inner sep=0pt, minimum size=5pt] (a2) at (-1,-0.5) [] {};
\node[circle, draw, inner sep=0pt, minimum size=5pt] (a3) at (-1,0) [] {};
\draw[dotted, thick] (-1,-.3) -- (-1,-.7);

\node[circle, draw, inner sep=0pt, minimum size=5pt] (b) at (-10,0) [label=below:$b$] {};
\node[circle, draw, inner sep=0pt, minimum size=5pt] (d) at (-11.5,-1.5) [label=below:$d$] {};
\node[circle, draw, inner sep=0pt, minimum size=5pt] (c) at (-10,-3) [label=below:$c$] {};

\node[circle, draw, inner sep=0pt, minimum size=5pt] (b1) at (-11,.4) [] {};
%\node[circle, draw, inner sep=0pt, minimum size=5pt] (b2) at (-11,0) [] {};
\node[circle, draw, inner sep=0pt, minimum size=5pt] (b3) at (-11,-.4) [] {};
\draw[dotted, thick] (-11,0.1) -- (-11,-0.1);

\node[circle, draw, inner sep=0pt, minimum size=5pt] (c1) at (-11,-3.4) [] {};
%\node[circle, draw, inner sep=0pt, minimum size=5pt] (c2) at (-11,-3) [] {};
\node[circle, draw, inner sep=0pt, minimum size=5pt] (c3) at (-11,-2.6) [] {};
\draw[dotted, thick] (-11,-2.9) -- (-11,-3.1);

%\node[circle, draw, inner sep=0pt, minimum size=5pt] (d1) at (-12.5,-.7) [] {};
\node[circle, draw, inner sep=0pt, minimum size=5pt] (d2) at (-12.5,-1.1) [] {};
\node[circle, draw, inner sep=0pt, minimum size=5pt] (d3) at (-12.5,-1.9) [] {};
\draw[dotted, thick] (-12.5,-1.4) -- (-12.5,-1.6);

\node (a00) at (-1,-0.5) [label=right:${V_a}$] {};
\node (a00) at (-12,0) [label=right:${V_b}$] {};
\node (a00) at (-12,-3) [label=right:${V_c}$] {};
\node (a00) at (-13.5,-1.5) [label=right:${V_d}$] {};
\node (a00) at (-1,-2.5) [label=right:${V_e}$] {};
\node () at (-6.5,-4.5) [label=right:$G'$] {};
\node () at (-16.5,-4.5) [label=right:$G$] {};
\path 
(v1) edge node[left]{$e_1$} (v2)
(v2) edge node[below]{$e_2$} (v3)
(v1) edge node[right]{$e_3$} (v3)
(a) edge (e1)
(a) edge (e2)
(a) edge (e3)
(e1) edge (v01)
(e1) edge (v02)
(e2) edge (v02)
(e2) edge (v03)
(e3) edge (v01)
(e3) edge (v03)
(v11) edge (v01)
(v12) edge (v02)
(v13) edge (v03)
(b) edge (v11)
(b) edge (v12)
(b) edge (v13)
(c) edge (v11)
(c) edge (v12)
(c) edge (v13)
(d) edge (b)
(d) edge (c)
(d) edge (a)
(a) edge (a1)
%(a) edge (a2)
(a) edge (a3)
(b) edge (b1)
%(b) edge (b2)
(b) edge (b3)
(c) edge (c1)
%(c) edge (c2)
(c) edge (c3)
%(d) edge (d1)
(d) edge (d2)
(d) edge (d3)
(d) edge (t)
(e1) edge (t)
(e2) edge (t)
(e3) edge (t)
%(t1) edge (t)
(t2) edge (t)
(t3) edge (t)
;
\end{tikzpicture}
\caption{An illustration of the reduction from {\sc Vertex Cover} to {\sc  Offensive Alliance} in Theorem \ref{ETHTH}. }
   \label{fig:ETH}
\end{figure}
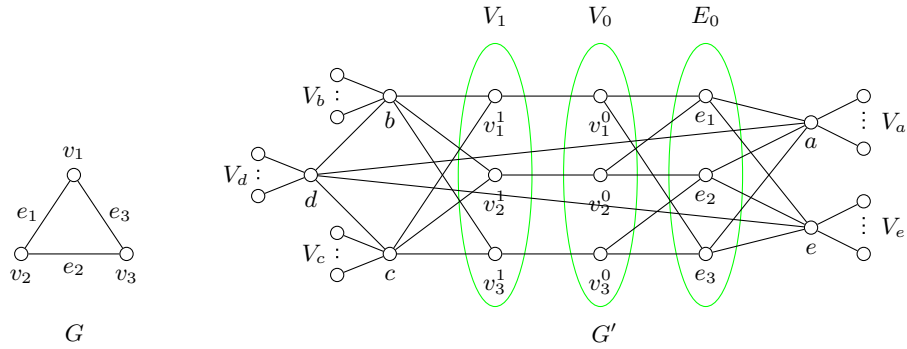

\begin{theorem}\label{ETHTH} \rm
 Unless ETH fails, {\sc  Offensive Alliance}  does not admit a $2^{o(n)}$ algorithm,
 even when restricted to bipartite graphs.
 \end{theorem}
\proof We give a linear reduction from {\sc Vertex Cover} in graphs of maximum degree 3  to {\sc Offensive Alliance}, that is, a 
polynomial-time algorithm that takes an instance  of {\sc Vertex Cover}  and 
outputs an equivalent instance of {\sc Offensive Alliance} whose size is bounded by $O(n)$.
Let $(G,k)$ be an instance of {\sc Vertex Cover}, where $G=(V,E)$ has maximum degree 3.
 We construct an equivalent instance $(G',k')$
of {\sc Offensive Alliance} the following way.  See Figure \ref{fig:ETH} for an illustration. 
Take two distinct copies $V_0,V_1$ of $V=\{v_1,v_2,\ldots,v_n\}$,  and let $v^i$ be the copy
of $v\in V$ in $V_i$. 
We introduce the  vertex set $E_0$ 
into $G'$, where $E_0=\{e_1,\ldots,e_m\}$, the edge set of $G$. 
We make $v^0_i$ adjacent to $e_j$ if and only if $v_i$ is an endpoint of $e_j$ in $G$.
We  make $v^0_{i}$ adjacent to $v^1_{i}$ for all $i$. 
Next, introduce five new vertices $a,b,c,d,e$.  For each $x\in \{a,b,c,d,e\}$, introduce a 
set $V_x$ of  $4k'$ vertices and make $x$ adjacent to every vertex of $V_x$. 
Moreover, we make vertex $a$ and $e$ adjacent to every vertex of  $E_0$ and make  $b$ and $c$ adjacent to every vertex of  $V_1$. We also make $d$ adjacent to every vertex of
$\{a,b,c,e\}$. 
Note that $G'$ is  a bipartite graph with bipartition 
$  \{d\} \cup V_1 \cup E_0 \bigcup\limits_{x\in \{a,b,c\}} {V_x}$ and $ \{a,b,c,e\} \cup V_d \cup V_0 $. We set $k'=k+5$. Clearly, the size 
of $G'$ is bounded by $O(n)$. 

We claim that $(G,k)$ is a yes-instance of {\sc Vertex Cover} if and only if 
$(G',k')$ is a yes-instance of 
{\sc Offensive Alliance}. 
Suppose $G$ has a vertex cover $S$ of 
size at most $k$. We show that $D=\{ v^0\in V_0~|~ v\in S \}\cup \{a,b,c,d,e\} $ is an 
 offensive alliance of size at most $k'$ in $G'$. We see  $N(D) = E_0 \cup V_1 
\bigcup\limits_{x\in\{a,b,c,d\}}{V_x}$. It is clear that for each 
$v\in V_1 \bigcup\limits_{x\in\{a,b,c,d\}}{V_x}$, 
we have $d_{D}(v)\geq d_{D^c}(v) +1$.
Each $v\in E_0$ has at least three neighbours in $D$, more precisely, $a$, $e$ and at least one neighbour in $V_0$.
This implies that for each $v \in E_0$, we have $d_{D}(v)\geq d_{D^c}(v) +1$.
\par Conversely, assume that $G'$ admits an offensive alliance $D$ of size at most $k'=k+5$.
Note that the vertices $a,b,c,d$ and $e$ cannot be part of the set $N(D)$ as 
each  of them has degree $4k'$; otherwise the 
size of $D$ will exceed $k'$. We now show that $\{a,b,c,d,e\}\subseteq D$. 
By definition,  offensive alliance cannot be empty; therefore it must contain a vertex from $V(G')$. 
\case  Suppose  $D$ contains  an arbitrary vertex of $\bigcup\limits_{x\in \{a,b,c,d,e\}}{V_x}$.
Without loss of generality, assume that $D$ contains an arbitrary vertex  of  $V_a$. 
Since $a \not\in N(D)$, it implies that $a\in D$. Since $a\in D$, 
we get $b,c,d$ and $e$ also lie in $D$ as otherwise $\{b,c,d,e\}\subseteq N(D)$.
Therefore, if any vertex from the set $\{a,b,c,d,e\}$ is in $D$, it implies that the whole set is in $D$. 
\case  Suppose $D$ contains an arbitrary vertex of $V_1$. 
It  implies that $b,c$ are in $D$ and therefore $\{a,b,c,d,e\}\subseteq D$.
\case Suppose $D$ contains  an arbitrary vertex of $E_0$. 
It implies that $\{a,e\}\subseteq D$ and therefore $\{a,b,c,d,e\}\subseteq D$.
\case Suppose $D$ contains $v^0$ from $V_0$.
This implies that  $v^1\in V_1$ is in $N(D)$. 
%(Note that if it is in $D$ then we can give the same argument as in Case 2). 
If both $b$ and $c$ are outside $D$ then we have $d_{D}(v^{1})<d_{D^c}(v^{1})+1$, 
which is a contradiction. This implies that either $b$ or $c$ is in $D$ and therefore $\{a,b,c,d,e\}\subseteq D$.

Now since $\{a,b,c,d,e\}\subseteq D$, a vertex $e$ in $E_0$ will be either in $N(D)$ or  $D$. 
If $e\in  D$, then we  pick an arbitrary neighbour of $e$ in $V_0$ and put it in $D$ and 
remove $e$ from $D$. Therefore, staring with an arbitrary offensive alliance $D$, 
we can transform it into another
 offensive alliance such that $D\cap E_0=\emptyset$, that is, $E_0\subseteq N(D)$. 
As each $e\in E_0$  has to satisfy the condition $d_S(e)\geq d_{S^c}(e)+1$, we must have a set $S\subseteq V_0$ of size at most $k$   in $D$ such that 
every vertex in $E_0$ 
has at least one neighbour in $S$. This implies that $S$ is a vertex cover of size at most $k$ in $G$. \qed

\subsection{ {\sc Offensive Alliance} on apex graphs}

\begin{theorem}\label{ETHapex}\rm
Unless ETH fails, the {\sc Strong Offensive Alliance} problem does not admit a $2^{o(\sqrt{n})}$ algorithm even when restricted to apex graphs.
\end{theorem}

\proof We give a linear reduction from {\sc Planar dominating set} to {\sc Strong Offensive Alliance}, that is, a 
polynomial-time algorithm that takes an instance of {\sc Planar dominating set} on $n$ vertices and $m=\mathcal{O}(n)$ edges, and 
outputs an equivalent instance of {\sc Strong Offensive Alliance} whose size is bounded by $\mathcal{O}(n)$.
Let $(G,k)$ be an instance of {\sc Planar dominating set}. Without loss of generality, we assume that $G$ is connected.  We construct an equivalent instance $(G',k')$
of {\sc Strong Offensive Alliance} in the following way. See Figure \ref{fig:ETH2} for an illustration. To construct graph $G'$, we start with graph $G$. Now for every edge $e\in E(G)$, we add one parallel edge $e'$ with same endpoints. Now, we take subdivision of newly added parallel edges and denote the edge vertex by $v_{e}$ corresponding to parallel edge of $e$.
Next, we make every edge vertex $v_{e}$ adjacent to three new vertices $\{{1}^{e},h_{2}^{e},h_{3}^{e}\}$.
We introduce two new vertices $x$ and $x'$. Finally, we add a set $V_{x}^{\square}$ of $6n$ vertices and make all of them adjacent to both $x$ and $x'$. 
Lastly, we make $x$ adjacent to all the vertices in the set $\bigcup\limits_{e\in E(G)} \{v_{e},h_{1}^{e},h_{2}^{e},h_{3}^{e}\}$. 
This completes the construction of $G'$. We set $k'=m+k+2$ We observe that deleting vertex $x$ makes the graph $G'$ planar. Therefore, $G'$ is an apex graph. \\
\par Formally, we claim that $G$ has a dominating set of size at most $k$ if and only if $G'$ has an strong offensive alliance of size at most $k'$.
Let us assume that $G$ admits a dominating set $D$ of size at most $k$.
We claim that $S=D \cup \{x,x'\} \bigcup\limits_{e\in E(G)} \{v_{e}\} $ is an strong offensive alliance of size at most $k'$.
It is easy to see that $|S|\leq k'$. First we observe that $N(S) = V(G')\setminus S \cup V_{x}^{\square}$. We see that for all vertices $v\in \bigcup\limits_{e\in E(G)} \{h_{1}^{e}, h_{2}^{e}, h_{3}^{e}\}$, we have $N_{S}(v) \geq N_{S^{c}}(v)+2$. 
It is also easy to see that same inequality is true for all the vertices in the set $V_{x}^{\square}$ as well. 
We observe that if a vertex $v \in V(G)$ have degree $d$ in $G$ then $d_{G'}(v) = 2d$. Since $D$ is a dominating set, the vertex $v$ has at least $d+1$ neighbours in $S$ and at most $d-1$ neighbours outside $S$. 
This implies that $N_{S}(v) \geq N_{S^{c}}(v)+2 $. Therefore $S$ is an strong offensive alliance. \\
\par In the reverse direction, suppose now that $S$ is an strong offensive alliance of size at most $k'$. We first show that $\{x,x'\}\subseteq S$.
As the degree of both $x$ and $x'$ is more than $2k'$, we conclude that $\{x,x'\}\cap N(S) = \emptyset$.
Therefore, if any vertex from the set $V_{x}^{\square} \bigcup\limits_{e\in E(G)} \{v_{e},h_{1}^{e}, h_{2}^{e}, h_{3}^{e}\}$ lie in $S$ then $x\in S$.
Now, we consider the last case where some original vertex $u\in V(G)$ in $G'$ lie in $S$. It implies that a vertex $v_{e}$ lies inside $S$ or $N(S)$ for some edge $e$ adjacent to $u$. The first case is trivial. For the second case, assume that $v_{e}\in N(S)$. Clearly, satisfying the inequality $N_{S}(v_{e}) \geq N_{S^{c}}(v_{e})+2$ requires at least one vertex from $\{x,h_{1}^{e}, h_{2}^{e}, h_{3}^{e}\}$ inside a solution. This shows that the vertex $x$ lies inside a solution $S$. It also implies that $x'\in S$. As the vertex $x\in S$, we have $ \bigcup\limits_{e\in E(G)} \{v_{e},h_{1}^{e}, h_{2}^{e},h_{3}^{e}\}\in S \cup N(S)$. First, we observe that if $v_{e}\in S$ for some $e$ then $S\setminus \{h_{1}^{e}, h_{2}^{e}, h_{3}^{e}\}$ is also a strong offensive alliance. Therefore we assume that if $v_{e}\in S$ then $ S \cap \{h_{1}^{e}, h_{2}^{e}, h_{3}^{e}\}= \emptyset$. We also observe that if $v_{e}\in N(S)$ then $ \{h_{1}^{e}, h_{2}^{e}, h_{3}^{e}\} \subseteq S$. We construct $S'$ from $S$ in a following way such that $\{x,x'\} \bigcup\limits_{e\in E(G)} \{v_{e}\} \in S'$. For every $v_{e}\in N(S)$ where $e=\{a,b\}$, we add $v_{e}$ to $S$ and the vertices $a$ and $b$ inside a solution and remove the set $\{h_{1}^{e}, h_{2}^{e}, h_{3}^{e}\}$ from a solution. It easy to see that $|S'|\leq |S|$. We will show that $S'$ is also a strong offensive alliance. First, we observe that $ \bigcup\limits_{e\in E(G)}\{h_{1}^{e}, h_{2}^{e}, h_{3}^{e}\} \subseteq N(S)$ and all of them satisfy $N_{S}(v) \geq N_{S^{c}}(v)+2$. Now, consider any vertex $v\in V(G)$ such that $v\in N(S')$. We consider the following two cases whether $v\in N(S)$  or $v \not\in N(S)$. In the first case, we observe that since we are only increasing the neighbours of $v$ inside $S'$, it will satisfy $N_{S}(v) \geq N_{S^{c}}(v)+2$. For the second case, we observe that if  $v \not\in N(S)$ but $v\in N(S')$ then it is only possible if we added a neighbour $u$ of $v$ from $V(G)$ inside $S'$. Let us assume that $d_{G}(v)=d$. Since $\bigcup\limits_{e\in E(G)} \{v_{e}\} \in S'$ it implies that $N_{S'}(v)\geq d+1$ and therefore $N_{S'}(v)\leq d-1$. This shows that $S'$ is a strong offensive alliance. We claim that $S' \cap V(G)$ is a dominating set of size at most $k$.  As we have already proved that $\{x,x'\} \cup \bigcup\limits_{e\in E(G)} \{v_{e}\} \in S'$, it implies that $|S' \cap V(G)|\leq k$. It is easy to see that $V(G) \subseteq (S'\cup N(S'))$. For all the vertices $v\in N(S')$, we have at least $d+1$ neighbours of $v$ inside $S'$. As $|\{v_{e} ~|~  \text{e is adjacent to v} \}|=d$ where $d=d_{G}(v)$, $v$ must have at least one neighbour from $S' \cap V(G)$. This proves that $D$ is a dominating set in $G$ of size at most $k$. \\

\begin{theorem}\label{Apex}
The {\sc Strong Offensive Alliance} problem admits a $\mathcal{O}^{*}(2^{\mathcal{O}{(\sqrt{n} \log{n})}})$ algorithm even when restricted to apex graphs.
\end{theorem}

Note that the treewidth of any apex graph with $n$ vertices is bounded by $\mathcal{O}(\sqrt{n})$. In \cite{ICDCIT2021}, a polynomial time algorithm is given
to solve offensive alliance problem on bounded treewidth graphs with running time $\mathcal{O}^{*}(2^\omega n^{\mathcal{O}(\omega)})$ where $\omega$ 
denotes the treewidth of the input graph. This algorithm can be used to obtain an 
algorithm with running time $\mathcal{O}^{*}(2^{\mathcal{O}{(\sqrt{n} \log{n})}})$
for apex graphs.\\

\section{NP-completeness results}
In this section, we prove that the {\sc Offensive Alliance} problem is NP-complete, even when restricted to split, chordal and circle graphs.

\subsection{Split and Chordal Graphs} A graph $G$ is called {\it chordal} if it does not contain any chordless cycle of length at least four. 
Split graphs are a subclass of chordal graphs, where the vertex set can be partitioned into an independent set and a clique.
We now prove the following theorem. 
\begin{theorem}\label{ETHTH10}
The {\sc Offensive Alliance} problem is NP-complete, even when restricted to split or chordal graphs.
\end{theorem}
 
\begin{figure}[ht]
    \centering
   \begin{tikzpicture}[scale=0.5]
%\draw [draw=green](-4,0) ellipse (.7cm and 2.6cm); 
%\draw [draw=green](-8,0) ellipse (.7cm and 2.6cm); 
%\draw [draw=green](-4,-4.2) ellipse (.6cm and 1.4cm); 
%\draw [draw=green](-8,-4.2) ellipse (.6cm and 1.5cm);
\draw [draw=blue](-1.8,-1.5) ellipse (2.8cm and 5cm); 
\draw [draw=red](-8.5,-1.5) ellipse (1.3cm and 5cm); 
\node[circle, draw, inner sep=0pt, minimum size=3pt] (v1) at (-16,-1.5) [label=above:${v_{1}}$] {};
\node[circle, draw, inner sep=0pt, minimum size=3pt] (v2) at (-17,-3) [label=below:${v_{2}}$] {};
\node[circle, draw, inner sep=0pt, minimum size=3pt] (v3) at (-15,-3) [label=below:${v_{3}}$] {};

\node (x1) at (-16.5,-2.3) [label=left:${e_{1}}$] {};
\node (x1) at (-15.5,-2.3) [label=right:${e_{2}}$] {};
\node (x1) at (-16,-3) [label=below:${e_{3}}$] {};

\node (x1) at (-16,-4.3) [label=below:${G}$] {};
\node (x1) at (-6,-7) [label=below:${G'}$] {};

\node[circle, draw, inner sep=0pt, minimum size=3pt] (v01) at (-1,1) [label=right:${e_{1}}$] {};
\node[circle, draw, inner sep=0pt, minimum size=3pt] (v02) at (-2.3,0) [label=right:${e_{2}}$] {};
\node[circle, draw, inner sep=0pt, minimum size=3pt] (v03) at (-3.6,-1) [label=right:${e_{3}}$] {};
\draw[green](v03)..controls(-2.3,0.5)..(v01);

\node[circle, draw, inner sep=0pt, minimum size=3pt] (e1) at (-8,1) [label=left:${v_{1}}$] {};
\node[circle, draw, inner sep=0pt, minimum size=3pt] (e2) at (-8,0) [label=left:${v_{2}}$] {};
\node[circle, draw, inner sep=0pt, minimum size=3pt] (e3) at (-8,-1) [label=left:${v_{3}}$] {};

\node[circle, draw, inner sep=0pt, minimum size=3pt] (z4) at (-1,-5) [label=right: $y_4$ ] {};
\node[circle, draw, inner sep=0pt, minimum size=3pt] (z3) at (-2,-4) [label=right: $y_3$] {};
\node[circle, draw, inner sep=0pt, minimum size=3pt] (z2) at (-3,-3) [label=right: $y_2$] {};
\node[circle, draw, inner sep=0pt, minimum size=3pt] (z1) at (-4,-2) [label=right: $y_1$] {};
\draw[green](z1)..controls(-3,-3.5)..(z3);
\draw[green](z2)..controls(-2,-4.5)..(z4);
\draw[green](z1)..controls(-3,-4.5)..(z4);
\path
(z1) edge [green] (z2)
(z2) edge [green] (z3)
(z3) edge [green] (z4);

\node[circle, draw, inner sep=0pt, minimum size=3pt] (h1) at (-8,-2) [label=left:${x_{1}}$] {};
\node[circle, draw, inner sep=0pt, minimum size=3pt] (h24) at (-8,-5) [label=left:${x_{24}}$] {};

\draw[dotted, thick] (-8,-3) --(-8,-4);

\draw (h1)--(z1);
\draw (h1)--(z2);
\draw (h1)--(z3);
\draw (h1)--(z4);
\draw (h24)--(z1);
\draw (h24)--(z2);
\draw (h24)--(z3);
\draw (h24)--(z4);
\draw(v1)--(v2);
\draw(v2)--(v3);
\draw(v3)--(v1);

\draw(v01)--(e1);
\draw(v01)--(e2);
\draw(v02)--(e3);
\draw(v02)--(e1);
\draw(v03)--(e2);
\draw(v03)--(e3);

\node (v00) at (-9.5,0) [label=left:$V_v$] {};
\node (v00) at (-9.5,-3.5) [label=left:$X$] {};

\node (v00) at (1,0) [label=right:$V_e$] {};
\node (v00) at (1,-4.2) [label=right:$Y$] {};
\path
(v01) edge [green] (z1)
(v01) edge [green] (z2)
(v01) edge [green] (z3)
(v01) edge [green] (z4)
(v02) edge [green] (z1)
(v02) edge [green] (z2)
(v02) edge [green] (z3)
(v02) edge [green] (z4)
(v03) edge [green] (z1)
(v03) edge [green] (z2)
(v03) edge [green] (z3)
(v03) edge [green] (z4)
(v01) edge [green] (v02)
(v02) edge [green] (v03);

\end{tikzpicture}
\caption{An illustration of the reduction from {\sc Vertex Cover} to {\sc  Offensive Alliance} in Theorem \ref{ETHTH10}. }
   \label{fig:ETH2}
\end{figure}
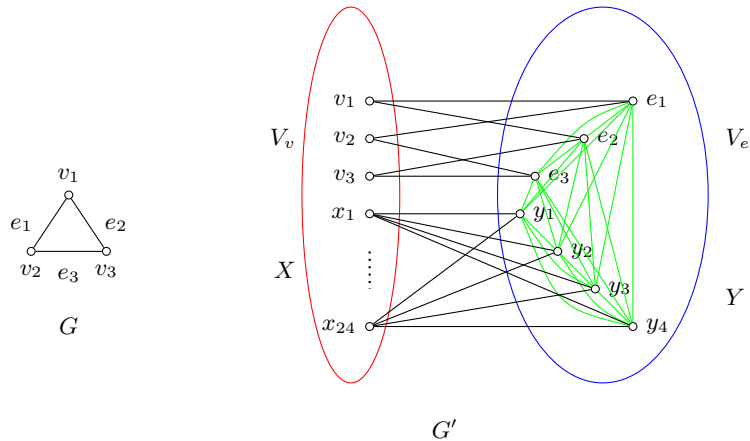

\proof It is easy to see  that the problem is in NP. 
To show that the problem is NP-hard we give a polynomial  reduction from {\sc Vertex Cover} in graphs of maximum degree 3.
Let $(G,k)$ be an instance of {\sc Vertex Cover}, where $G$ has maximum degree 3. We construct an equivalent instance $(G',k')$
of {\sc Offensive Alliance}  the following way. See Figure \ref{fig:ETH2} for an illustration. 

\noindent We construct  $G^{\prime}$ with  vertex sets $V$ and $V_e$, where
$V=\{v_1,v_2,\ldots,v_n\}$ and $V_e=E(G)=\{e_1,e_2,\ldots,e_m\}$, the edge set of $G$. 
We make $v_i$ adjacent to $e_j$ if and only if $v_i$ is an endpoint of $e_j$. 
Make all the vertices of $V_e$ pairwise adjacent.
We further add a set $Y=\{y_1, y_2,\ldots,y_{m+1}\}$ of $m+1$ vertices; 
the vertices in $Y$ are adjacent to every element of $V_e$ and the vertices 
in $Y$ are pairwise adjacent. Note that the vertices of $V_e\cup Y$ 
form a clique of size $2m+1$.
Finally we introduce a set $X=\{x_1,\ldots,x_{4(n+m)}\}$ of $4(n+m)$ vertices.
We make every vertex of $X$ adjacent to every vertex of  $Y$. 
Note  that $V\cup X$ forms an independent set where as the 
vertices in $V_e\cup Y$ form a clique. 
Therefore, $G$ is a split graph. 
We set $k'=k+m+1$. 
%As $m=O(n)$, the number of vertices in $G'$ is bounded by $O(n)$ and the number of edges in $G'$ is bounded by $O(n^2)$. \\

Formally, we claim that $G$ has a vertex cover of size at most $k$ if and
only if $G'$ has an offensive alliance of size at most $k'$. Assume first that
$G$ admits a vertex cover $S$ of size at most $k$. Consider $ D = S \cup Y $.
Clearly, $|D|\leq k'$.
 We claim that $D$ is an offensive alliance in $G'$. Note that $N(D) = V_e \cup X $. 
 For each  $x\in X$, we have $d_D(x)\geq d_{D^c}(x)+1$ 
 as all its neighbours are inside $D$. Each  $e \in V_e$ has
 at least $m+2$ neighbours in $D$ and at most $m+1$ neighbours, 
 including itself,  outside $D$. 
 This implies that   $D$ is an offensive alliance of size at most $k'$ in $G'$. 
 
\par For the reverse direction, let $D$ be an offensive alliance of size at most $k'$  in  $G'$. 
We first show that $Y \subseteq D$. It is easy to note that $Y \cap N(D) = \emptyset$ as
otherwise each $v\in Y \cap N(D)$ has to satisfy the condition $d_D(v)\geq d_{D^c}(v)+1$
which requires more than $k'$ vertices in $D$. 
Since $D$ is a non-empty offensive alliance, it must contain a vertex from the 
set $V \cup V_e \cup X \cup Y$.\\

\noindent{\it Case 1:} Suppose $D$ contains a vertex  from $ V_e \cup X \cup Y$. 
 As  $Y \cap N(D) = \emptyset$, therefore we get  $Y \subseteq D$.\\

\noindent{\it Case 2:}  Suppose $D$ contains a vertex $v$ from $V$. Let
$e\in V_e$ be a neighbour of $v$ in $G'$. Then $e$ could be either in $D$ or in $N(D)$.
If $e$ is in $D$, then Case 1 implies that $Y \subseteq D$. 
Suppose $e$ is in $N(D)$. Then $e$ has to satisfy the condition 
$d_D(e) \geq d_{D^c}(e)+1$, which requires at least one 
vertex from $Y$ inside $D$. Again Case 1 implies that $Y \subseteq D$.\\

Since the size of $D$ is at most $m+k+1$ and $Y\subseteq D$, 
it can contain at most $k$ vertices besides the vertices in $Y$. 
Given an offensive alliance $D$, we can construct another offensive alliance $D'$ 
such that $|D'|\leq |D|$ and $D' \cap (V_e \cup X) = \emptyset$, in the 
following way. For each $e \in V_e \cap D$, we replace $e$ by an arbitrary 
neighbour of $e$ in $V_v$. If a neighbour of $e$ is already present in $D$ then 
just remove $e$ and do not add any new vertex. 
We also remove all the vertices of $X$ from $D$. 
The modified $D$ is our $D'$. 
Next we argue that $D'$ is an offensive alliance. 
Since $Y \subseteq D'$ and $D' \cap (V_e \cup X) = \emptyset$, 
we have $N(D') =V_e \cup X$. 
It is easy to see that each  $x\in X$ satisfies the condition
$d_{D'}(x)\geq d_{D'^c}(x)+1$.  
We know that  $V_e \subseteq D \cup N(D)$. 
We observe that for the vertices in set $N(D)\cap V_e$, we only increase their
number of neighbours in $D'$. Therefore, the vertices $v$ in  $N(D)\cap V_e$ 
satisfy the condition $d_{D'}(v)\geq d_{D'^c}(v)+1$. We see that for the vertices in $V_e \cap D$, 
we have at least one neighbour from $V$ inside $D'$ by 
the construction of $D'$. Clearly, each vertex in $V_e \cap D$ has
at least $m+2$ neighbours inside $D'$ and at most $m+1$ 
(including itself) outside $D'$. 
This shows that $D'$ is an offensive alliance. We also know that $V_e \subseteq  N(D')$ 
and therefore each vertex $e\in V_e$ has to satisfy the condition $d_{D'}(e)\geq d_{D'^c}(e)+1$
which requires at least one neighbour from $V$ inside  $D'$. 
Therefore $D' \cap V$ forms a vertex cover of size at most $k$ in $G$. 
This proves that $(G,k)$ is a yes instance.  \qed\\

%\noindent As a consequence of the above theorems, we get the following result:
%\begin{corollary}\label{ETHTH0}
%The {\sc Offensive Alliance} problem is NP-complete even when restricted to bipartite, chordal or split graphs.
%\end{corollary}

\subsection{Circle graphs}

%A {\it circle graph} is the intersection graph of a set of chords of a circle. 
A {\it circle graph} is an undirected graph whose vertices can be associated with chords of a circle such that two vertices are adjacent if and only if the corresponding chords cross each other.
%A number of  problems that are NP-complete on general graphs have polynomial time algorithms when restricted to circle graphs. For instance, Kloks \cite{KLOKS-1996} showed that the treewidth of a circle graph can be determined, and an optimal tree decomposition constructed, in $O(n^3)$ time;  Tiskin \cite{Tiskin}  has shown that a maximum clique of a circle graph can be found in $O(n~ log^2 n)$ time. 
Here, we prove that the {\sc Offensive Alliance} problem is NP-complete even when restricted to circle graphs, via a reduction from {\sc Dominating Set}.  It is known  that the {\sc Dominating Set} problem  on circle graphs is NP-hard \cite{KEIL199351}.

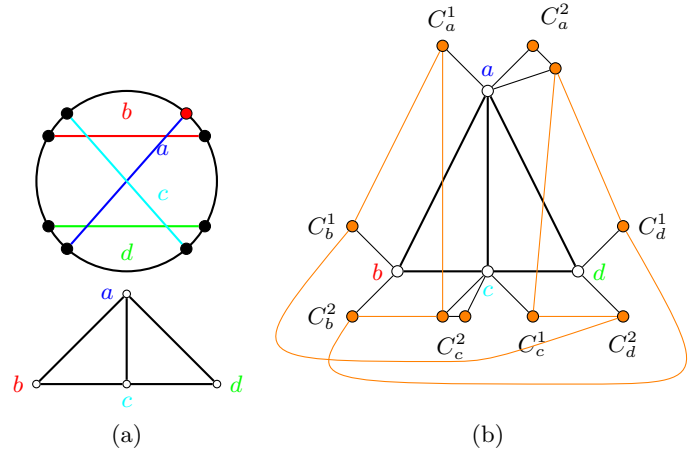
\begin{figure}
\begin{center}
\begin{tikzpicture}[scale=0.6]

\draw[thick] (0,0) circle [radius=2];

\node[fill=black, circle, draw=black, inner sep=0, minimum size=0.15cm](q1) at (1.732,1) [] {};
\node[fill=black, circle, draw=black, inner sep=0, minimum size=0.15cm](q2) at (-1.732,1) [] {};
\node(k1) at (0,1) [label=above:$\color{red} { b}$] {};
\node(k1) at (0.8,0.2) [label=above:$\color{blue} {a}$] {};
\node(k1) at (0,-1) [label=below:$\color{green} {d}$] {};
\node(k1) at (0.8,-0.8) [label=above:$\color{cyan} {c}$] {};
\node[fill=black, circle, draw=black, inner sep=0, minimum size=0.15cm](q3) at (-1.732,-1) [] {};
\node[fill=black, circle, draw=black, inner sep=0, minimum size=0.15cm](q4) at (1.732,-1) [] {};

\node[fill=red, circle, draw=black, inner sep=0, minimum size=0.15cm](q5) at (1.322, 1.5) [] {};
\node[fill=black, circle, draw=black, inner sep=0, minimum size=0.15cm](q6) at (1.322,-1.5) [] {};
\node[fill=black, circle, draw=black, inner sep=0, minimum size=0.15cm](q7) at (-1.322, 1.5) [] {};
\node[fill=black, circle, draw=black, inner sep=0, minimum size=0.15cm](q8) at (-1.322, -1.5) [] {};

\draw[thick, red](q1)--(q2);
\draw[thick,  green](q3)--(q4);
\draw[thick, blue](q5)--(q8);
\draw[thick, cyan](q6)--(q7);

\node[circle, draw=black, inner sep=0, minimum size=0.1cm](a0) at (0,-2.5) [label=left:$\color{blue} {a}$] {};
\node[circle, draw=black,inner sep=0, minimum size=0.1cm](c0) at (0,-4.5) [label=below:$\color{cyan} {c}$] {};
\node(k1) at (0,-5) [label=below:(a)] {};
\node(k1) at (8,-5) [label=below:(b)] {};
\node[circle, draw=black, inner sep=0,minimum size=0.1cm](b0) at (-2,-4.5) [label=left:$\color{red} { b}$] {};
\node[circle, draw=black, inner sep=0,minimum size=0.1cm](d0) at (2,-4.5) [label=right:$\color{green} {d}$] {};

\node[circle, draw=black, inner sep=0, minimum size=0.15cm](a) at (8,2) [label=above:$\color{blue} {a}$] {};

\node[circle, fill=orange, draw=black, inner sep=0, minimum size=0.15cm](a1) at (7,3) [label=above:$C_a^1$] {};

\node(k1) at (9.5,3) [label=above:$C_a^2$] {};

\node[circle, fill=orange, draw=black, inner sep=0, minimum size=0.15cm](a2) at (9,3) [] {};
\node[circle, fill=orange, draw=black, inner sep=0, minimum size=0.15cm](a3) at (9.5,2.5) [] {};

\node[circle, draw=black,inner sep=0, minimum size=0.15cm](c) at (8,-2) [label=below:$\color{cyan} {c}$] {};

\node[circle, fill=orange, draw=black,inner sep=0, minimum size=0.15cm](c1) at (7,-3) [] {};
\node[circle, fill=orange, draw=black,inner sep=0, minimum size=0.15cm](c2) at (7.5,-3) [] {};
\node[circle, fill=orange, draw=black,inner sep=0, minimum size=0.15cm](c3) at (9,-3) [label=below:$C_c^1$] {};

\node(kc) at (7.2,-3) [label=below:$C_c^2$] {};

\node[circle, draw=black, inner sep=0,minimum size=0.15cm](b) at (6,-2) [label=left:$\color{red} { b}$] {};

\node[circle, fill=orange, draw=black, inner sep=0,minimum size=0.15cm](b1) at (5,-1) [label=left:$C_b^1$] {};
\node[circle, fill=orange, draw=black, inner sep=0,minimum size=0.15cm](b2) at (5,-3) [label=left:$C_b^2$] {};

\node[circle, draw=black, inner sep=0,minimum size=0.15cm](d) at (10,-2) [label=right:$\color{green} {d}$] {};

\node[circle, fill=orange, draw=black, inner sep=0,minimum size=0.15cm](d1) at (11,-1) [label=right:$C_d^1$] {};
\node[circle, fill=orange, draw=black, inner sep=0,minimum size=0.15cm](d2) at (11,-3) [label=below:$C_d^2$] {};

\draw(a)--(a1);
\draw(a)--(a2);
\draw(a)--(a3);
\draw(c)--(c1);
\draw(c)--(c2);
\draw(c)--(c3);
\draw(b)--(b1);
\draw(b)--(b2);
\draw(d)--(d1);
\draw(d)--(d2);
\draw(a2)--(a3);
\draw(c1)--(c2);
\draw[thick](a)--(b);
\draw[thick](a)--(c);
\draw[thick](a)--(d);
\draw[thick](b)--(c);
\draw[thick](d)--(c);
\draw[orange](a1)--(b1);
\draw[thick](a0)--(b0);
\draw[thick](a0)--(c0);
\draw[thick](a0)--(d0);
\draw[thick](b0)--(c0);
\draw[thick](d0)--(c0);
\draw[orange](b2)--(c1);
\draw[orange](a1)--(c1);
\draw[orange](c3)--(a3);
\draw[orange](d2)--(c3);

\draw[orange](a3)--(d1);

\draw[orange] (b1) .. controls(2.5,-4) .. (7,-4)..controls(8,-4).. (d2);

\draw[orange] (d1) .. controls(13,-4.5) .. (10,-4.5)..controls(4,-4.5).. (b2);

\end{tikzpicture}
\caption{ (a) Graph $G$ and its circle representation. (b) The graph $G'$ produced by the 
reduction algorithm. Note that every orange vertex  is adjacent to a set of $2r$
vertices, which are not shown here.}
\label{OAcircle}
\end{center}
 \end{figure}

 \begin{theorem}\label{FNvds}\rm
The {\sc Offensive Alliance} problem on circle graphs is NP-complete.
 \end{theorem}
\proof It is easy to see  that the problem is in NP. 
To show that the problem is NP-hard we give a polynomial reduction from 
{\sc Dominating Set} on circle graphs. Let $(G,k)$ be an instance of {\sc Dominating Set},
where $G$ is a circle graph. 
Suppose we are also given the circle  representation $C$ of  $G$. 
Without loss of generality, we assume that there are no one degree vertices in $G$.
We construct an instance of $(G',r)$ of {\sc Offensive Alliance} as follows (see
Figure \ref{OAcircle}). Set $r=2m+k$, where $m$ is the number of edges in $G$.
For every $v\in V(G)$, we introduce two cliques $C_v^1$ and $C_v^2$ where
$C_v^1$ has $\lfloor\frac{d(v)}{2}\rfloor$ nodes 
and $C_v^2$ has $\lceil\frac{d(v)}{2}\rceil$ nodes; make $v$ adjacent to every vertex 
of $C_v^1\cup C_v^2$; for 
every $x\in C_v^1\cup C_v^2$,
 introduce  a set $V_{v,x}^{\square}$ of $2r$ new vertices and make
 $x$ adjacent to every vertex of $V_{v,x}^{\square}$. 
We start at an arbitrary
 vertex of the circle representation $C$ of $G$ and then traverse the circle in a clockwise direction. 
 We record the sequence in which the chords are visited. 
 For example, in Figure \ref{OAcircle}, if we start at
 the red vertex on the circle, 
 then the sequence in which the chords are visited, is $a,b,d,c,a,d,b,c$.
 Note that every vertex  appears twice in the sequence as every chord 
 is visited twice while traversing the circle. Thus we get a sequence $S$ of 
 length $2n$ where
 $n$ is the number of chords in $C$. We use the sequence to connect $2n$ newly added
 cliques.  For every consecutive pair $(u,v)$ in the sequence $S$, put
 an edge between a vertex of $C_u^1$ (resp. $C_u^2$) and a vertex 
 of $C_v^1$ (resp. $C_v^2)$ if both $u,v$  appear for the first time (resp. second time)
 in the sequence;  put
 an edge between a vertex of $C_u^1$ and a vertex 
 of $C_v^2$ if $u$ appears for the first time
 and $v$ appears for the second time in the sequence.
   These edges are shown in orange  in Figure \ref{OAcircle}.  
  This completes the construction of graph $G'$. 
 Now we show that $G'$ is  indeed a circle graph. 
 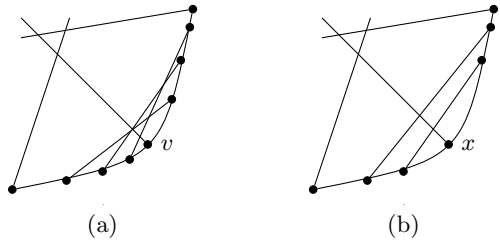
\begin{figure}
     \centering
 \begin{tikzpicture}[scale=0.4]
 
  \node (x0) at (6,6) [] {};
   \node (x00) at (16,6) [] {};
   
\node (x000) at (8,6) [] {};
\node (x0000) at (6,5) [] {};

\node (y000) at (18,6) [] {};
\node (y0000) at (16,5) [] {};
 
 \node[fill=black, circle, draw=black, inner sep=0, minimum size=0.1cm](y1) at (6,0) [label=below:] {};
 
 \node[fill=black, circle, draw=black, inner sep=0, minimum size=0cm] at (9,-0.5) [label=below:(a)] {}; 
\node[fill=black, circle, draw=black, inner sep=0, minimum size=0cm] at (19,-0.5) [label=below:(b)] {}; 
\node[fill=black, circle, draw=black, inner sep=0,minimum size=0.1cm](y2) at (12,6) [label=right:] {};

\node[fill=black, circle, draw=black, inner sep=0,minimum size=0.1cm](y3) at (10.5,1.5) [label=right:$v$] {};

\draw (y1)..controls(11,1)..(y2);

\node[fill=black, circle, draw=black, inner sep=0,minimum size=0.1cm](w1) at (11.9,5.4) [] {};

\node[fill=black, circle, draw=black, inner sep=0,minimum size=0.1cm](w2) at (11.6,4.3) [] {};

\node[fill=black, circle, draw=black, inner sep=0,minimum size=0.1cm](w3) at (11.3,3) [] {};

\node[fill=black, circle, draw=black, inner sep=0,minimum size=0.1cm](w4) at (9.9,1) [] {};

\node[fill=black, circle, draw=black, inner sep=0,minimum size=0.1cm](w5) at (9,0.6) [] {};

\node[fill=black, circle, draw=black, inner sep=0,minimum size=0.1cm](w6) at (7.8,0.3) [] {};

 \node[fill=black, circle, draw=black, inner sep=0,minimum size=0.1cm](z1) at (16,0) [label=below:] {};

\node[fill=black, circle, draw=black, inner sep=0,minimum size=0.1cm](z2) at (22,6) [label=right:] {};

\node[fill=black, circle, draw=black, inner sep=0,minimum size=0.1cm](z3) at (20.5,1.5) [label=right:$x$] {};

\node[fill=black, circle, draw=black, inner sep=0,minimum size=0.1cm](w7) at (21.9,5.4) [] {};

\node[fill=black, circle, draw=black, inner sep=0,minimum size=0.1cm](w8) at (21.6,4.3) [] {};

%\node[fill=black, circle, draw=black, inner sep=0,minimum size=0.1cm](w9) at (21.3,3) [] {};

%\node[fill=black, circle, draw=black, inner sep=0,minimum size=0.1cm](w10) at (19.9,1) [] {};

\node[fill=black, circle, draw=black, inner sep=0,minimum size=0.1cm](w11) at (19,0.6) [] {};

\node[fill=black, circle, draw=black, inner sep=0,minimum size=0.1cm](w12) at (17.8,0.3) [] {};

\draw (z1)..controls(21,1)..(z2);

\draw(w1)--(w4);
\draw(w2)--(w5);
\draw(w3)--(w6);
\draw(w7)--(w12);
\draw(w8)--(w11);
%\draw(w9)--(w10);

\draw(y3)--(x0);
\draw(z3)--(x00);

\draw(y1)--(x000);
\draw(y2)--(x0000);
\draw(z1)--(y000);
\draw(z2)--(y0000);

\end{tikzpicture}

     \caption{(a) The circle representation for the first operation. Let $d(v)=6$. For $v$, we introduce $C_v^1$ in $G'$, and make $v$ adjacent with every vertex of 
     $C_v^1$. The circle representation of $v$, $C_v^1$ and their adjacency are shown here. 
     (b) The circle representation for the second operation with $r=1$.  }
     \label{fig:circlegadget}
 \end{figure}

\begin{figure}
     \centering
\begin{tikzpicture}[scale=0.4]

\draw[thick] (0,0) circle [radius=4];

\node(b) at (0,1.8) [label=above:$\color{red} {b}$] {};
\node(a) at (1,0.6) [label=above:$\color{blue} {a}$] {};
\node(c) at (1,-0.8) [label=above:$\color{cyan} {c}$] {};
\node(d) at (0,-1.8) [label=below:$\color{green} {d}$] {};

\node[fill=black, circle, draw=black, inner sep=0,minimum size=0.1cm](b1) at (3.6,1.7) [] {};
\node[fill=black, circle, draw=black, inner sep=0,minimum size=0.1cm](b2) at (-3.6,1.7) [] {};
\node[fill=black, circle, draw=black, inner sep=0,minimum size=0.1cm](d1) at (3.6,-1.7) [] {};
\node[fill=black, circle, draw=black, inner sep=0,minimum size=0.1cm](d2) at (-3.6,-1.7) [] {};

\node[fill=black, circle, draw=black, inner sep=0,minimum size=0.1cm](a1) at (2,3.5) [] {};
\node[fill=black, circle, draw=black, inner sep=0,minimum size=0.1cm](c1) at (-2,3.5) [] {};
\node[fill=black, circle, draw=black, inner sep=0,minimum size=0.1cm](c2) at (2,-3.5) [] {};
\node[fill=black, circle, draw=black, inner sep=0,minimum size=0.1cm](a2) at (-2,-3.5) [] {};

\node[fill=black, circle, draw=black, inner sep=0,minimum size=0.1cm](x1) at (-3.35,2.2) [] {};
\node[fill=black, circle, draw=black, inner sep=0,minimum size=0.1cm](x2) at (-3,2.6) [] {};
\node[fill=black, circle, draw=black, inner sep=0,minimum size=0.1cm](x3) at (-2.6,3) [] {};

\node[fill=black, circle, draw=black, inner sep=0,minimum size=0.1cm](x4) at (3.9,0.8) [] {};
\node[fill=black, circle, draw=black, inner sep=0,minimum size=0.1cm](x5) at (-3.9,0.8) [] {};
\node[fill=black, circle, draw=black, inner sep=0,minimum size=0.1cm](x6) at (3.9,-0.8) [] {};
\node[fill=black, circle, draw=black, inner sep=0,minimum size=0.1cm](x7) at (-3.9,-0.8) [] {};

\node[fill=black, circle, draw=black, inner sep=0,minimum size=0.1cm](x8) at (-3.35,-2.2) [] {};
\node[fill=black, circle, draw=black, inner sep=0,minimum size=0.1cm](x9) at (-3,-2.6) [] {};
\node[fill=black, circle, draw=black, inner sep=0,minimum size=0.1cm](x10) at (-2.6,-3) [] {};

\node[fill=black, circle, draw=black, inner sep=0,minimum size=0.1cm](x11) at (0,4) [] {};
\node[fill=black, circle, draw=black, inner sep=0,minimum size=0.1cm](x12) at (-1,3.85) [] {};
\node[fill=black, circle, draw=black, inner sep=0,minimum size=0.1cm](x13) at (1,3.85) [] {};

\node[fill=black, circle, draw=black, inner sep=0,minimum size=0.1cm](x14) at (0,-4) [] {};
\node[fill=black, circle, draw=black, inner sep=0,minimum size=0.1cm](x15) at (-1,-3.85) [] {};
\node[fill=black, circle, draw=black, inner sep=0,minimum size=0.1cm](x16) at (1,-3.85) [] {};

\node[fill=black, circle, draw=black, inner sep=0,minimum size=0.1cm](x17) at (3.35,2.2) [] {};
\node[fill=black, circle, draw=black, inner sep=0,minimum size=0.1cm](x18) at (2.6,3) [] {};

\node[fill=black, circle, draw=black, inner sep=0,minimum size=0.1cm](x19) at (3.35,-2.2) [] {};
\node[fill=black, circle, draw=black, inner sep=0,minimum size=0.1cm](x20) at (2.6,-3) [] {};

\draw(x1)--(x12);
\draw(x2)--(x7);
\draw(x3)--(x13);
\draw(x18)--(x6);
\draw(x11)--(x17);
\draw(x4)--(x20);
\draw(x19)--(x14);
\draw(x16)--(x10);
\draw(x15)--(x8);
\draw(x9)--(x5);

\draw[thick, red] (b1)--(b2);
\draw[thick, green] (d1)--(d2);
\draw[thick, blue] (a1)--(a2);
\draw[thick, cyan] (c1)--(c2);

\end{tikzpicture}
     \caption{A circle representation of the graph $G'$ in Figure \ref{OAcircle}. We do not shown the 
     parallel chords correspond to  $2r$ vertices adjacent to every vertex in each clique.}
     \label{fig:democircle}
 \end{figure}
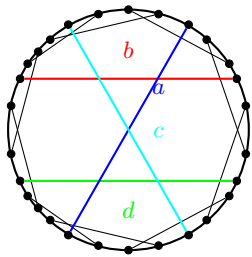

\noindent In the reduction algorithm, we have three operations: (i) For every $v\in V(G)$, 
 we introduce two cliques $C_v^1$
and $C_v^2$ and make $v$ adjacent to every vertex 
of $C_v^1$ and $C_v^2$.
This operation can be incorporated in the circle representation 
by introducing $\lfloor\frac{d(v)}{2}\rfloor $ intersecting chords 
at  one end of the chord corresponds to $v$ and $\lceil\frac{d(v)}{2}\rceil$ intersecting 
chords at the other end of the chord corresponds to $v$. See Figure \ref{fig:circlegadget} for an illustration.
 (ii) For every vertex $x$ in clique, we introduce  a set of $2r$ new vertices and make
 $x$ adjacent to each of them. This operation can be easily incorporated in the 
 circle representation by introducing 
 $2r$ parallel chords intersecting the chord corresponds to $x$.  See Figure \ref{fig:circlegadget}.
 (iii) For every consecutive pair $(u,v)$ in the sequence $S$, we put
 an edge between a vertex of $C_u^1$ and a vertex 
 of $C_v^1$. This is incorporated in the circle representation by 
 making the last chord (in clockwise direction) of $C_u^1$ intersect with 
 the first chord (in clockwise direction)  of $C_v^1$. This is demonstrated in Figure \ref{fig:democircle}.
 
 \par Formally, we claim that $G$ has a dominating set of size at most $k$ if and
only if $G'$ has an offensive alliance of size at most $r$. Assume first that 
$G$ admits a dominating set $S$ of size at most $k$.
 Consider $$ D = \bigcup\limits_{v\in V(G)} V(C_v^1) 
 \cup V(C_v^2)   \cup S. $$ Clearly $|D|\leq r$.
 We claim that $D$ is an offensive alliance  in $G'$. Clearly 
 $$N(D) =  (V(G)\setminus S)  \bigcup\limits_{v\in V(G)}
 \bigcup\limits_{x \in C_v^1\cup C_v^2} 
 V_{v,x}^{\square}.$$
 It is easy to see that each $u\in \bigcup\limits_{v\in V(G)}
 \bigcup\limits_{x \in C_v^1\cup C_v^2} 
 V_{v,x}^{\square}$ satisfies $d_D(u)\geq d_{D^c}(u)+1$. For  $u \in V(G)\setminus S$, if $d_{G}(u)=d$ 
 then $d_{D}(u)\geq d+1$ and $d_{D^c}(u)\leq d-1$ in $G'$. 
 Thus $D$ is an offensive alliance of size at most $r$ in $G'$.

\par Conversely,  suppose $G'$ admits an  offensive alliance $D$ of size at most $r$. 
 First, we show that $$\bigcup\limits_{v\in V(G)} V(C_v^1) 
 \cup V(C_v^2) \subseteq D.$$ 
 It is to be noted that any offensive alliance $D$ of size at most $r$ 
 cannot contain a vertex of degree more than $2r$ in its neighbourhood $N(D)$. 
 This implies that 
$ N(D) \bigcap  \bigcup\limits_{v\in V(G)} V(C_v^1) 
 \cup V(C_v^2)  = \emptyset$. 
 Since $D$ is a non-empty offensive alliance, 
 it must contain a vertex from  $G'$. Suppose $D$ contains a vertex from 
 $\bigcup\limits_{v\in V(G)} V(C_v^1) 
 \cup V(C_v^2)$.
 Since $N(D)$ cannot contain an element of $\bigcup\limits_{v\in V(G)} V(C_v^1) 
 \cup V(C_v^2)$, 
 it implies that $\bigcup\limits_{v\in V(G)} V(C_v^1) 
 \cup V(C_v^2) \subseteq D$. Suppose $D$ contains
 an element from 
 $V(G) \cup \bigcup\limits_{v\in V(G)}
 \bigcup\limits_{x \in C_v^1\cup C_v^2} 
 V_{v,x}^{\square}$ then it is clear from the above argument that  
 $\bigcup\limits_{v\in V(G)} V(C_v^1)\cup V(C_v^2) \subseteq D$. It is to be noted that 
 the total number of vertices in $\bigcup\limits_{v\in V(G)} V(C_v^1) 
 \cup V(C_v^2)$ is $\sum\limits_{v\in V}d(v)=2m$, where
 $m$ is the number of edges in $G$. 
 We have added $2m$ vertices inside the solution. Apart from these vertices, we can  add $k$ more vertices inside the solution. We observe that the vertices in $V(G)$ are either in $D$ or $N(D)$. 
 Each $v\in V(G) \cap N(D)$ needs exactly one neighbour from $V(G)\cap D$, in addition
 to its neighbours in $C_v^1
 \cup C_v^2$,
 in order to satisfy the condition $d_{D}(v)\geq d_{D^c}(v)+1$. 
 Therefore,  $D \cap V(G)$ is a dominating set of size at most $k$ in $G$. \qed

 \section{Conclusion and and Future Directions} 
 In this work we proved that the {\sc Offensive Alliance} problem is NP-complete even when restricted to bipartite, chordal, split and circle graphs. We proved that 
 the {\sc Offensive Alliance} problem is W[1]-hard
 parameterized by a wide range of 
fairly restrictive structural parameters such as the feedback vertex set number, treewidth, pathwidth, and treedepth of the input graph  thus not FPT (unless FPT = W[1]).
We thereby resolved an open question stated by Bernhard Bliem  and Stefan Woltran  (2018) concerning the complexity of 
{\sc Offensive Alliance} parameterized by treewidth.
 This is especially interesting because most ``subset problems'' that are FPT when parameterized by solution size turned out to be FPT for the parameter treewidth \cite{DomMichael}, and moreover {\sc Offensive Alliance} is easy on trees. We gave lower bound based on ETH
for the time needed to solve the {\sc Offensive Alliance} problem; we proved that it cannot be solved in time $2^{o(n)}$ even when restricted to bipartite graphs, unless ETH fails.  
We list some natural questions that arise from the results of this study:
\begin{itemize}
    \item Does {\sc Offensive Alliance}  parameterized by vertex cover number admit a single exponential algorithm or can one show a lower bound with matching time complexity?
    \item Does {\sc Offensive Alliance} admits polynomial-time algorithms on some  special classes of intersection graph family such as interval graphs, circular arc graphs, unit disk graphs, etc?
    \item Determine parameterized complexity of {\sc Offensive Alliance} problem when parameterized by other structural parameters such as twin cover, cluster vertex deletion number and modular-width.
   % \item Finally, it may be interesting to study if our ideas can be useful for different kinds of alliances from the literature such as defensive alliances, powerful alliances and secure set.
\end{itemize}

\bibliographystyle{abbrv}
\bibliography{bibliography}

\begin{thebibliography}{10}

\bibitem{Basavaraju2018OnTK}
M.~Basavaraju, F.~Panolan, A.~Rai, M.~S. Ramanujan, and S.~Saurabh.
\newblock On the kernelization complexity of string problems.
\newblock {\em Theor. Comput. Sci.}, 730:21--31, 2018.

\bibitem{BLIEM2018334}
B.~Bliem and S.~Woltran.
\newblock Defensive alliances in graphs of bounded treewidth.
\newblock {\em Discrete Applied Mathematics}, 251:334 -- 339, 2018.

\bibitem{CHANG2012479}
C.-W. Chang, M.-L. Chia, C.-J. Hsu, D.~Kuo, L.-L. Lai, and F.-H. Wang.
\newblock Global defensive alliances of trees and cartesian product of paths
  and cycles.
\newblock {\em Discrete Applied Mathematics}, 160(4):479 -- 487, 2012.

\bibitem{chel}
M.~Chellali and T.~W. Haynes.
\newblock Global alliances and independence in trees.
\newblock {\em Discuss. Math. Graph Theory}, 27(1):19--27, 2007.

\bibitem{marekcygan}
M.~Cygan, F.~V. Fomin, L.~Kowalik, D.~Lokshtanov, D.~Marx, M.~Pilipczuk,
  M.~Pilipczuk, and S.~Saurabh.
\newblock {\em Parameterized Algorithms}.
\newblock Springer, 2015.

\bibitem{DomMichael}
M.~Dom, D.~Lokshtanov, S.~Saurabh, and Y.~Villanger.
\newblock Capacitated domination and covering: A parameterized perspective.
\newblock In M.~Grohe and R.~Niedermeier, editors, {\em Parameterized and Exact
  Computation}, pages 78--90, Berlin, Heidelberg, 2008. Springer Berlin
  Heidelberg.

\bibitem{Downey}
R.~G. Downey and M.~R. Fellows.
\newblock {\em Parameterized Complexity}.
\newblock Springer, 2012.

\bibitem{Enciso2009AlliancesIG}
R.~Enciso.
\newblock {\em Alliances in graphs: Parameterized algorithms and on
  partitioning series -parallel graphs}.
\newblock PhD thesis, University of Central Florida, USA, 2009.

\bibitem{fellows}
M.~R. Fellows, D.~Lokshtanov, N.~Misra, F.~A. Rosamond, and S.~Saurabh.
\newblock Graph layout problems parameterized by vertex cover.
\newblock In S.-H. Hong, H.~Nagamochi, and T.~Fukunaga, editors, {\em
  Algorithms and Computation}, pages 294--305, Berlin, Heidelberg, 2008.
  Springer Berlin Heidelberg.

\bibitem{Fernau}
H.~Fernau and D.~Raible.
\newblock Alliances in graphs: a complexity-theoretic study.
\newblock In {\em Proceeding Volume II of the 33rd International Conference on
  Current Trends in Theory and Practice of Computer Science}, 2007.

\bibitem{DA-FPT}
H.~Fernau and D.~Raible.
\newblock Alliances in graphs: a complexity-theoretic study.
\newblock In J.~van Leeuwen, G.~F. Italiano, W.~van~der Hoek, C.~Meinel,
  H.~Sack, F.~Plasil, and M.~Bielikov{\'{a}}, editors, {\em {SOFSEM} 2007:
  Theory and Practice of Computer Science, 33rd Conference on Current Trends in
  Theory and Practice of Computer Science, Harrachov, Czech Republic, January
  20-26, 2007, Proceedings Volume {II}}, pages 61--70. Institute of Computer
  Science {AS} CR, Prague, 2007.

\bibitem{FERNAU2009177}
H.~Fernau, J.~A. Rodríguez, and J.~M. Sigarreta.
\newblock Offensive $r$-alliances in graphs.
\newblock {\em Discrete Applied Mathematics}, 157(1):177 -- 182, 2009.

\bibitem{frick}
G.~Fricke, L.~Lawson, T.~Haynes, M.~Hedetniemi, and S.~Hedetniemi.
\newblock A note on defensive alliances in graphs.
\newblock {\em Bulletin of the Institute of Combinatorics and its
  Applications}, 38:37--41, 2003.

\bibitem{ICDCIT2021}
A.~Gaikwad, S.~Maity, and S.~K. Tripathi.
\newblock Parameterized complexity of defensive and offensive alliances in
  graphs.
\newblock In D.~Goswami and T.~A. Hoang, editors, {\em Distributed Computing
  and Internet Technology}, pages 175--187, Cham, 2021. Springer International
  Publishing.

\bibitem{mss}
R.~Ganian, F.~Klute, and S.~Ordyniak.
\newblock On structural parameterizations of the bounded-degree vertex deletion
  problem.
\newblock {\em Algorithmica}, 2020.

\bibitem{Paturi}
R.~Impagliazzo, R.~Paturi, and F.~Zane.
\newblock Which problems have strongly exponential complexity?
\newblock {\em Journal of Computer and System Sciences}, 63(4):512--530, 2001.

\bibitem{10.5555/1292785}
L.~H. Jamieson.
\newblock {\em Algorithms and Complexity for Alliances and Weighted Alliances
  of Various Types}.
\newblock PhD thesis, Clemson University, USA, 2007.

\bibitem{Lindsay}
L.~H. Jamieson, S.~T. Hedetniemi, and A.~A. McRae.
\newblock The algorithmic complexity of alliances in graphs.
\newblock {\em Journal of Combinatorial Mathematics and Combinatorial
  Computing}, 68:137--150, 2009.

\bibitem{soda1999}
D.~S. Johnson and M.~Szegedy.
\newblock What are the least tractable instances of max independent set?
\newblock In {\em Proceedings of the Tenth Annual ACM-SIAM Symposium on
  Discrete Algorithms}, SODA '99, page 927–928, USA, 1999. Society for
  Industrial and Applied Mathematics.

\bibitem{KEIL199351}
J.~Keil.
\newblock The complexity of domination problems in circle graphs.
\newblock {\em Discrete Applied Mathematics}, 42(1):51--63, 1993.

\bibitem{KIYOMI201791}
M.~Kiyomi and Y.~Otachi.
\newblock Alliances in graphs of bounded clique-width.
\newblock {\em Discrete Applied Mathematics}, 223:91 -- 97, 2017.

\bibitem{Kloks94}
T.~Kloks.
\newblock {\em Treewidth, Computations and Approximations}, volume 842 of {\em
  Lecture Notes in Computer Science}.
\newblock Springer, 1994.

\bibitem{kris}
P.~Kristiansen, M.~Hedetniemi, and S.~Hedetniemi.
\newblock Alliances in graphs.
\newblock {\em Journal of Combinatorial Mathematics and Combinatorial
  Computing}, 48:157--177, 2004.

\bibitem{Daniel-2018}
D.~Lokshtanov, D.~Marx, and S.~Saurabh.
\newblock Slightly superexponential parameterized problems.
\newblock {\em SIAM Journal on Computing}, 47(3):675--702, 2018.

\bibitem{Sparsity}
J.~Nesetril and P.~O. de~Mendez.
\newblock {\em Sparsity: Graphs, Structures, and Algorithms}.
\newblock Springer Publishing Company, Incorporated, 2014.

\bibitem{Neil}
N.~Robertson and P.~Seymour.
\newblock Graph minors. iii. planar tree-width.
\newblock {\em Journal of Combinatorial Theory, Series B}, 36(1):49 -- 64,
  1984.

\bibitem{ROD}
J.~Rodríguez-Velázquez and J.~Sigarreta.
\newblock Global offensive alliances in graphs.
\newblock {\em Electronic Notes in Discrete Mathematics}, 25:157 -- 164, 2006.

\bibitem{SIGARRETA20091687}
J.~Sigarreta, S.~Bermudo, and H.~Fernau.
\newblock On the complement graph and defensive k-alliances.
\newblock {\em Discrete Applied Mathematics}, 157(8):1687 -- 1695, 2009.

\bibitem{SIGARRETA20061345}
J.~Sigarreta and J.~Rodríguez.
\newblock On defensive alliances and line graphs.
\newblock {\em Applied Mathematics Letters}, 19(12):1345 -- 1350, 2006.

\bibitem{SIGA}
J.~Sigarreta and J.~Rodríguez.
\newblock On the global offensive alliance number of a graph.
\newblock {\em Discrete Applied Mathematics}, 157(2):219 -- 226, 2009.

\bibitem{west}
D.~B. West.
\newblock {\em Introduction to Graph Theory}.
\newblock Prentice Hall, 2000.

\end{thebibliography}

\end{document}